\documentclass{article}

\usepackage{microtype}
\usepackage{graphicx}
\usepackage{subfigure}
\usepackage{pdfcomment}
\usepackage{booktabs} 

\usepackage{hyperref}

\usepackage[accepted]{icml2024}

\usepackage{amsmath}
\usepackage{amssymb}
\usepackage{mathtools}
\usepackage{amsthm}
\usepackage{tablefootnote}

\usepackage[capitalize,noabbrev]{cleveref}

\usepackage{microtype}
\usepackage{graphicx}
\usepackage{subfigure}
\usepackage{subcaption}
\usepackage{booktabs}
\usepackage{hyperref}
\usepackage{xspace}
\usepackage{amsmath}
\usepackage{mathtools}
\usepackage{amsfonts}
\PassOptionsToPackage{hyphens}{url}
\usepackage{enumitem}
\usepackage{algorithm}
\usepackage{algorithmic}

\let\oldnl\nl
\newcommand\nonl{%
  \renewcommand{\nl}{\let\nl\oldnl}}

\newcommand{\sys}{\textsc{HexGen}\xspace}

\newcommand{\x}{\mathbf{x}}

\newcommand{\wq}{\mathbf{w}_Q}
\newcommand{\wv}{\mathbf{w}_V}
\newcommand{\wk}{\mathbf{w}_K}
\newcommand{\wo}{\mathbf{w}_O}
\newcommand{\wa}{\mathbf{w}_1}
\newcommand{\wb}{\mathbf{w}_2}
\newcommand{\xq}{\mathbf{x}_Q}
\newcommand{\xv}{\mathbf{x}_V}
\newcommand{\xk}{\mathbf{x}_K}

\newcommand{\xo}{\mathbf{x}_{\text{Out}}}

\newcommand{\tx}{\mathbf{t}}
\newcommand{\tq}{\mathbf{t}_Q}

\newcommand{\txo}{\mathbf{t}_{\text{Out}}}

\newcommand{\rebuttal}[1]{\textcolor{black}{#1}}

\theoremstyle{plain}

\theoremstyle{definition}

\theoremstyle{remark}

\usepackage[textsize=tiny]{todonotes}

\icmltitlerunning{\sys: Generative Inference of \textcolor{black}{Large Language} Model over Heterogeneous Environment}

\begin{document}

\twocolumn[
\icmltitle{\sys: Generative Inference of \textcolor{black}{Large Language Model \\over Heterogeneous Environment}}



\icmlsetsymbol{equal}{*}

\begin{icmlauthorlist}
\icmlauthor{Youhe Jiang}{equal,yyy}
\icmlauthor{Ran Yan}{equal,yyy}
\icmlauthor{Xiaozhe Yao}{equal,comp}
\icmlauthor{Yang Zhou}{sch}
\icmlauthor{Beidi Chen}{sch}
\icmlauthor{Binhang Yuan}{yyy}
\end{icmlauthorlist}

\icmlaffiliation{yyy}{Department of Computer Science and Engineering, The Hong Kong University of Science and Technology, Hong Kong, China}
\icmlaffiliation{comp}{Department of Computer Science, ETH Zurich, Zürich, Switzerland}
\icmlaffiliation{sch}{Department of Electrical and Computer Engineering, Carnegie Mellon University, Pittsburgh, Pennsylvania}

\icmlcorrespondingauthor{Binhang Yuan}{biyuan@ust.hk}

\icmlkeywords{Machine Learning, ICML}

\vskip 0.3in
]



\printAffiliationsAndNotice{\icmlEqualContribution} 

\begin{abstract}
Serving generative inference of the \textcolor{black}{large language} model is a crucial component of contemporary AI applications.
This paper focuses on deploying such services in a heterogeneous and \textcolor{black}{cross-datacenter} setting to mitigate the substantial inference costs typically associated with \textcolor{black}{a single centralized datacenter}.
Towards this end, we propose \sys, a \textit{flexible distributed inference engine} that uniquely supports the asymmetric partition of generative inference computations over both tensor model parallelism and pipeline parallelism and allows for effective deployment across diverse GPUs interconnected by a fully heterogeneous network.
We further propose a \textit{sophisticated scheduling algorithm} grounded in constrained optimization that can adaptively assign asymmetric inference computation across the GPUs to fulfill inference requests while maintaining acceptable latency levels.
We conduct an extensive evaluation to verify the efficiency of \sys by serving the state-of-the-art \textsc{Llama-2 (70B)} model. The results suggest that \sys can choose to achieve up to $2.3\times$ lower latency deadlines or tolerate up to $4\times$ more request rates compared with the homogeneous baseline given the same budget. Our implementation is available at \url{https://github.com/Relaxed-System-Lab/HexGen}.
\end{abstract}

\vspace{-1.em}
\section{Introduction}


\textcolor{black}{Large language} models are distinguished by the vast scale of parameters being trained over a substantial pre-train corpus. Such extensive training enables them to be remarkably adaptable across a broad spectrum of downstream tasks~\cite{bommasani2021opportunities}.
In fact, \textcolor{black}{large language} models such as GPT-4~\cite{bubeck2023sparks}, Llama2-70B~\cite{touvron2023llama}, and Falcon-180B~\cite{falcon180b} have essentially revolutionized the way AI systems are developed and deployed, which have nourished a large number of advanced applications. 
In such an ecosystem, serving the generative inference requests for \textcolor{black}{large language} models presents a critical challenge --- given the unprecedented model scale, unlike classic machine learning models, parallel inference strategies have to be leveraged to accommodate the high computational and memory demands while ensuring low-latency generative inference outcomes.

The state-of-the-art inference service of the \textcolor{black}{large language} model is usually hosted in a \textcolor{black}{single} centralized data center with homogeneous high-performance GPUs, which can be very expensive in terms of the cloud service fee. The high cost of such deployment potentially limits the democratization of this great technique. Alternatively, the deployment of the \textcolor{black}{large language} model inference over a heterogeneous \textcolor{black}{cross-datacenter} environment can be a promising direction to reduce the inference cost, which has not been fully explored. 
The heterogeneous environment for foundation model inference service can encompass a wide range of options, including more affordable cloud services (such as spot instances~\cite{thorpe2023bamboo, athlur2022varuna} and serverless computing~\cite{guo2022hydrozoa}) to even fully decentralized platforms~\cite{yuan2022decentralized, borzunov2023petals} that leverage a diverse set of GPUs contributed by volunteers \textcolor{black}{in an extreme setting}.


However, deploying \textcolor{black}{large language} model inference across a heterogeneous environment presents some unique challenges. 
Unlike traditional machine learning models, \textcolor{black}{large language} model inference consists of two different phases: a prompt phase that handles a sequence of input tokens at once and a decoding phase where output tokens are generated step-by-step. Additionally, \textcolor{black}{large language} models require the adoption of specialized \textit{parallel inference strategies} to effectively distribute the intensive computations across multiple GPUs. The two most commonly employed approaches are \textit{tensor model parallelism} and \textit{pipeline parallelism}.
In terms of coordinating such a complicated distributed computation, there are some fundamental challenges stemming from the \textit{heterogeneity}:


\begin{itemize}[topsep=5pt, leftmargin=*]
    \vspace{-1.em}
    \item \textbf{Heterogeneous GPU computation capacity.} To fully leverage the economic GPU computation power, it is essential to employ a range of GPU types, each has distinct peak FLOPS, GPU device memory bandwidth, and GPU device memory constraints. However, most (if not all) distributed implementations of \textcolor{black}{large language} model inference frameworks do \textit{not accommodate this GPU diversity}, as they typically assume a homogeneous GPU cluster configuration. This results in an implementation that enforces complete symmetry in the distribution of the inference computation, e.g., requiring every pipeline stage to adhere to the same tensor model parallel degree, which essentially limits the performances of the distributed inference over heterogeneous GPUs. 
    
    \vspace{-0.5em}
    \item \textbf{Heterogeneous GPU connection.} The heterogeneity of the cross-GPU connection is even more significant. In a standard homogeneous setting, the intra-machine GPU connections usually rely on the same NVLink or PCIe, and the inter-machine GPU connections are often based on RDMA. While in a fully heterogeneous setting, the connections between each pair of GPUs can vary significantly, including both fast NVLink or PCIe connections and cross-geo-region slow connections. This heterogeneity of connections forces a scheduling algorithm to consider \textit{exponentially many more distinct allocation strategies} when compared with the homogeneous setting. 

    \vspace{-0.5em}
\end{itemize}


\vspace{-0.25em}
In order to overcome these challenges, we propose \sys, a \textcolor{black}{large language} model inference system that coordinates distributed inference computation over a set of GPUs with different computation capacities connected by heterogeneous connections. We provide flexible support of asymmetric parallel execution within the scope of tensor model parallelism and pipeline parallelism to accommodate the heterogeneous GPU computation capacity. We also propose a scheduling algorithm to determine an efficient allocation of inference computation under diversified connections. 
Concretely, we make the following contributions:

\vspace{-0.25em}
\textbf{\underline{Contribution 1.}} We implement the \sys system with the support of asymmetric tensor model parallelism and pipeline parallelism. Essentially, \sys allows each pipeline parallel stage to be assigned with a different number of layers and tensor model parallel degrees to flexibly accommodate the heterogeneity of different GPUs and fully unleash the potential of heterogeneous GPU powers.   

\vspace{-0.25em}
\textbf{\underline{Contribution 2.}}  We formally define the scheduling problem of serving the inference of multiple copies of the same foundation model concurrently over a set of heterogeneous GPU devices as a constrained optimization problem.
We concretely define computation cost, communication cost, and memory limits for such inference workflow. We then propose a two-phase optimization solution, where we use a dynamic programming algorithm to define the optimal layer of each pipeline, and a heuristic-based evolutionary algorithm to efficiently search for the optimal layout.

\vspace{-0.25em}
\textbf{\underline{Contribution 3.}} We evaluate \sys by conducting a comprehensive empirical study to compare the system and economic efficiency between the heterogeneous setting enabled by \sys and the standard homogeneous setting within a centralized data center serving the state-of-the-art \textsc{Llama-2 (70B)} model. We show that given the \textit{same} budget in terms of cloud service fees, \sys can choose to achieve up to $2.3\times$ lower latency deadlines or tolerate up to $4\times$ more traffic request rate compared with the homogeneous baseline. Additionally, when given only \textit{half} of the budget, \sys can still maintain a similar level of inference service compared to the homogeneous baseline.

\vspace{-0.25em}
\underline{\textbf{Overview.}} The rest of the paper is organized as follows. We
provide some preliminaries in Section \ref{sec:preliminary}; introduce our asymmetric parallel implementation for the generative inference in Section \ref{sec:asysmetric} and scheduling algorithm in Section \ref{sec:schedule}; present the experimental results in Section \ref{sec:eval}, summarize related work in Section \ref{sec:rel}, and conclude in Section \ref{sec:conclude}.



\vspace{-0.5em}
\section{Preliminary}
\label{sec:preliminary}

\textbf{Generative inference computation.}
A typical generative inference task of the transformer-based foundation model consists of two stages: i) the \textit{prefill} stage that takes a prompt sequence to compute the key-value cache (KV cache) for each transformer layer of the \textcolor{black}{large language} model; and ii) the \textit{decoding} stage which utilizes the previous KV cache to generate new tokens step-by-step and appends the new KV cache. The inference computation can be summarized below.
Denote the batch size by $b$, the prompt sequence length by $s^\text{in}$, the output sequence length by $s^\text{out}$, the hidden dimension of the transformer by $H$, and the total number of transformer layers by $L$. Given the weight matrices of a transformer layer specified by $\wk^\kappa, \wq^\kappa, \wv^\kappa, \wo^\kappa \in \mathcal{R}^{H \times H} $, $\wa^\kappa \in \mathcal{R}^{H \times 4H}$, and $\wb^\kappa \in \mathcal{R}^{4H \times H}$.
During the prefill phase, the input of the $\kappa$-th layer is specified by $\x^\kappa$, and \texttt{key}, \texttt{value}, \texttt{query}, and \texttt{output} of the attention layer is specified $\xk^\kappa, \xv^\kappa, \xq^\kappa, \xo^\kappa \in \mathcal{R}^{b \times s \times H}$. First, the cached \texttt{key}, \texttt{value} can be computed by:
\begin{equation*}
\begin{small}
\begin{array}{cc}
     & \xk^\kappa = \x^\kappa \cdot \wk^\kappa; \quad  \xv^\kappa = \x^\kappa \cdot \wv^\kappa
\end{array}
\end{small}
\vspace{-0.5em}
\end{equation*}
The rest of the computation in the $\kappa$-th layer is:
\begin{equation*}
\begin{small}
\begin{array}{cc}
& \xq^\kappa = \x^\kappa \cdot \wq^\kappa \\
& \xo^\kappa = f_{\text{softmax}}\left( \frac{\xq^\kappa {\xk^\kappa}^T}{\sqrt{H}}\right )\cdot \xv^\kappa \cdot \wo^\kappa + \x^\kappa \\
& \x^{\kappa+1} = f_{\text{relu}}\left(\xo^\kappa \cdot \wa^\kappa \right) \cdot \wb^\kappa + \xo^\kappa
\end{array}
\end{small}
\vspace{-0.5em}
\end{equation*}
During the decode phase, given $\tx^\kappa \in \mathcal{R}^{b \times 1 \times H}$ as the embedding of the current generated token in the $\kappa$-th layer, the inference computation needs to i) update the KV cache:
\begin{equation*}
\begin{small}
      \xk^\kappa \leftarrow \text{Concat}\left( \xk^\kappa, \tx^\kappa \cdot \wk^\kappa \right);\:
      \xv^\kappa \leftarrow \text{Concat}\left( \xv^\kappa, \tx^\kappa \cdot \wv^\kappa \right)
\end{small}
\vspace{-0.5em}
\end{equation*}
And ii) compute the output of the current layer:
\begin{equation*}
\begin{small}
\begin{array}{cc}
     & \tq^\kappa = \tx^\kappa \cdot \wq^\kappa \\
     & \txo^\kappa = f_{\text{softmax}}\left( \frac{\tq^\kappa {\xk^\kappa}^T}{\sqrt{H}}\right)\cdot \xv^\kappa \cdot \wo^\kappa + \tx^\kappa\\
     & \tx^{\kappa+1} = f_{\text{relu}}\left(\txo^\kappa\cdot \wa^\kappa \right) \cdot \wb^\kappa + \txo^\kappa
\end{array}
\end{small}
\end{equation*}

\vspace{-0.5em}
\textbf{Parallelism in generative inference.}
Given the scale of the state-of-the-art \textcolor{black}{large language} models, distributed or parallel execution is necessary. Two standard parallel strategies are usually included in inference: \textit{Pipeline parallelism}~\cite{huang2019gpipe,narayanan2019pipedream} partitions the foundation model into multiple stages and serves the inference computation as a pipeline, where each GPU or (group of GPUs) handles a stage. During the inference computation, the GPU(s) serving stage-($j$) needs to \texttt{send} the activations to the GPU(s) serving stage-($j{+}1$). For inference computation, pipeline parallelism cannot reduce the completion time for a \textit{single} request since only one stage can be active. However, the communication volume included in pipeline parallelism is much less when compared with tensor model parallelism, which is beneficial for slow GPU connections. 
\textit{Tensor model parallelism}~\cite{narayanan2021efficient} partitions the inference computation at the level of transformer layers over multiple GPUs, where the weight matrices are distributed both row-wisely and column-wisely. Two \texttt{AllReduce} operations are required to aggregate each layer's output activations. Tensor model parallelism splits both the data scan and computation among a tensor model parallel group, which can effectively scale out the inference computation if the connection is fast among the group. However, when the intra-group communication is not extremely fast (i.e., not by NVLink or PCIe), tensor model parallelism can perform poorly. 

\vspace{-0.5em}
\section{Asymmetric Parallel Implementation}
\label{sec:asysmetric}

The current foundation model service systems (such as huggingface Accelerate~\cite{huggingfaceAccelerate}, and FastTransformer~\cite{fastertransformer}) can only support a symmetric inference setup --- it has to set all the tensor model parallel groups to share the same degree for each pipeline stage serving the same number of transformer layers. We first illustrate a case study about why asymmetric parallel support is necessary under heterogeneous environments and enumerate our system implementation of this functionality.

\vspace{-0.5em}
\subsection{Case Study: Parallelism over Heterogeneity}
\begin{figure}[t]
    \centering
    \includegraphics[width=\linewidth]{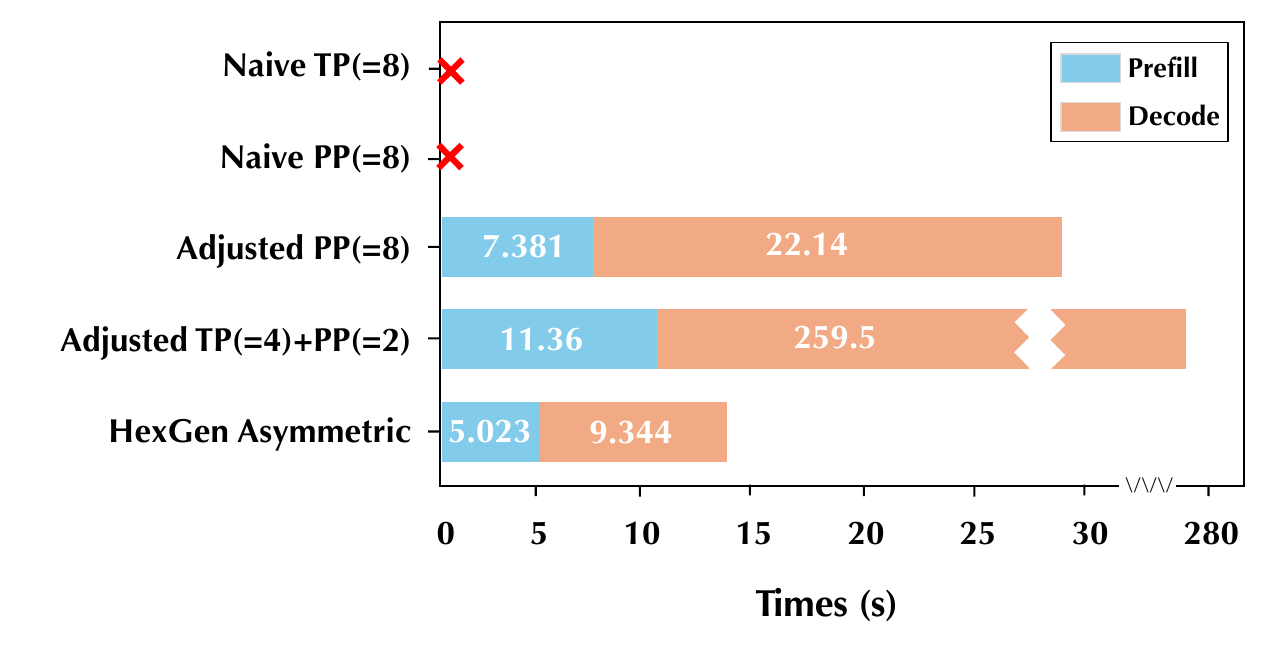}
    \vspace{-2em}
    \caption{Case study of parallel strategy over heterogeneity.}
    \label{fig:enter-label}
    \vspace{-2em}
\end{figure}

\vspace{-0.5em}
Consider a heterogeneous environment where we can use three GPU instances to serve a \textsc{Llama-2 (70B)} model: the first instance has \texttt{4$\times$A6000-48G}, the second instance has \texttt{2$\times$A5000-24G}, and the third instance has \texttt{2$\times$A4000-16G}. We test a generative inference request with an input length of $128$, and output length of $64$. The results are shown in Figure \ref{fig:enter-label}. We have some interesting observations: \underline{First}, \textit{direct usage of pure tensor model parallel parallelism (TP) or pipeline parallelism (PP) will cause the out-of-memory (OOM) error}: for tensor model parallelism, \texttt{A4000-16G} cannot host the evenly partitioned parameter shards; for naive pipeline parallelism, one need to evenly partition the transformer layers, where \texttt{A4000-16G} cannot hold $10$ layers of \textsc{Llama-2 (70B)} model on its stage. \underline{Second}, \textit{simply partitioning pipeline stage according to computation capacity leads to poor performance}: we test two layouts: i) set PP degree to $8$, and partition layer number proportionally with the GPU capacity, in this case, since the pipeline is long where only one stage can be active for computation, leading to slow inference; ii) set PP degree to $2$, and TP degree to $4$, where we use the first instance \texttt{4$\times$A6000-48G} to serve the first stage handling $56$ layers and the second and third instances \texttt{2$\times$A5000-24G$+$2$\times$A4000-16G} to handle the rest $24$ layers, in this case, the tensor model parallelism introduces significant cross-machine communication that compromises the performance. \underline{Lastly}, \textit{fully asymmetric allocation unleashes the performance under heterogeneity}: to avoid the issue we encounter above, we show the performance of the \sys, where we let each machine serve a single stage and handle $48$, $20$, $12$ layers with TP degrees as $4$, $2$, $2$, respectively --- we observe $2\times$ and $19\times$ speedup compared with the symmetric parallel executions. 

\vspace{-0.5em}
\subsection{\sys Asymmetric Parallel Implementation}
\vspace{-0.5em}
In order to implement the asymmetric parallel support, our essential change can be summarized as follows: \textit{each pipeline parallel stage can be assigned with a different number of layers and tensor model parallel degree}. From the perspective of the implementation, we make the following changes: i) when initializing the pipeline parallel communication group, \sys will take the configured tensor model parallel degree for each stage along with their corresponding number of allocated transformer layers; ii) each stage will select a leader GPU which introduces lowest communication latency to GPUs in the nearby stages, and initialize an independent tensor model parallel group; iii) when executing the inference through the pipeline, only the leader \rebuttal{GPU} in each stage (i.e., tensor model parallel group) will \texttt{send} the activation to the leader GPU in the next stage; once the leader GPU \texttt{receive}s the activation, it will \texttt{broadcast} this activation among its tensor model parallel group to execute the tensor model parallel computation. In terms of system implementation, we modify the latest FlashAttention~\cite{dao2023flashattention} framework by integrating this new pipeline parallel design to enable this functionality.

\begin{table*}[t!]
\caption{Modeling the generative inference cost and limit.}
\begin{small}
\label{tab:formula}
\vspace{-1em}
\begin{center}
\resizebox{\textwidth}{!}{%
\begin{tabular}{c | c | c}
\hline
\textbf{Description}  & \textbf{Notation} & \textbf{Cost Formulation} \\
\hline
Computation cost & 
$\textsc{C}^{i,j}_{\text{comp}}\left(\mathbf{d}_{i,j}\right)$&
\label{eq:comp_i_j}
$
\begin{aligned}
\max_{d \in \mathbf{d}_{i,j}}\left( \frac{12H^2 B_{\text{type}}s_t^{\text{out}}}{\left|\mathbf{d}_{i,j}\right| m_d} \right) \cdot l_{i,j} + & \max_{d \in \mathbf{d}_{i,j}}\left( \frac{24 b_t \left(s^{\text{in}}_t + s^{\text{out}}_t\right) H^2}{\left|\mathbf{d}_{i,j}\right| c_d}  \right) \cdot l_{i,j} 
\end{aligned}
$ \\  \hline
TP communication cost
&
$\textsc{C}^{i,j}_{\text{comm-tp}}\left(\mathbf{d}_{i,j}\right)$
&
$
\begin{aligned}
\max_{d \in \mathbf{d}_{i,j}}\left( \sum_{d' \in \mathbf{d}_{i,j} - \{d\}} \left(\alpha_{d, d'} + \frac{b_t s^{\text{in}}_{t} H B_{\text{type}} } { {\left|\mathbf{d}_{i,j}\right| \beta}_{d, d'}}\right) \right) \cdot 4 l_{i,j} + \max_{d \in \mathbf{d}_{i,j}} \left( \sum_{d' \in \mathbf{d}_{i,j} - \{d\}} \left(\alpha_{d, d'} + \frac{b_t H B_{\text{type}} } { {\left|\mathbf{d}_{i,j}\right| \beta}_{d, d'}}\right)  \right) \cdot 4 s_{t}^{\text{out}} l_{i,j}
\end{aligned}
$ \\ \hline
PP communication cost
&
$\textsc{C}^{i,j}_{\text{comm-pp}}\left(\mathbf{d}_{i,j}\right)$
&
$
\begin{aligned}
\min_{d \in \mathbf{d}_{i,j}, d' \in \mathbf{d}_{i,j+1}}\left( \alpha_{d,d'} + \frac{b_t s^{\text{in}}_{t} H B_{\text{type}}}{\beta_{d,d'}} \right) + \min_{d \in \mathbf{d}_{i,j}, d' \in \mathbf{d}_{i,j+1}}\left( \alpha_{d,d'} + \frac{b_t H B_{\text{type}}}{\beta_{d,d'}} \right) \cdot s_{t}^{\text{out}} 
\end{aligned}
$ \\ \hline
Memory limit
&
$\textsc{C}^{d}_{\text{mem}}\left(\mathbf{d}_{i,j}\right) $
&
$
\begin{aligned}
\left(\frac{12H^2 B_{\text{type}}}{\left|\mathbf{d}_{i,j}\right|} + \frac{2 b_t \left(s^{\text{in}}_t + s^{\text{out}}_t\right) H B_{\text{type}}}{\left|\mathbf{d}_{i,j}\right|} \right) \times l_{i,j} + \quad 4 b_t \left(s^{\text{in}}_t + s^{\text{out}}_t\right) H B_{\text{type}}
\end{aligned}
$ \\ 
\hline
\end{tabular}%
}
\end{center}
\end{small}
\scriptsize{We formulate computation cost, tensor model parallel (TP) communication cost, memory limit of the $j$-th stage in the $i$-th pipeline, and the pipeline parallel (PP) communication cost between the $j$-th stage and the $j{+}1$-th stage of the $i$-th pipeline for a particular inference task $t \in \mathbf{T}$, where $b_{t}$ is the batch size, $s^{\text{in}}_{t}$ is the sequence length of input prompt and $s^{\text{out}}_{t}$ is the number of output tokens, and $B_{\text{type}}$ denotes the number of bytes for the precision of inference computation, e.g., $B_{\text{type}}\left(\textsc{fp16}\right)=2$. (Details in Appendix \ref{sec:cost_model})} 
\label{tab:notations}
\end{table*}

\section{Scheduling over Heterogeneity}
\label{sec:schedule}
We introduce our scheduling algorithm in this Section.

\vspace{-0.5em}
\subsection{Problem Formalization}
\label{sec: problem}
\vspace{-0.5em}


\noindent \textbf{Notations.} We first introduce the following notations: Let $\mathbf{D} = \{d_1 \ldots d_N\}$ be a set of $N$ GPU devices, where $M_d$ denotes GPU memory limit, $m_d$ denotes GPU memory bandwidth, and $c_d$ denotes tensor core computation power; $\mathbf{A} \in \mathbb{R}_+^{N\times N}$ and $\mathbf{B} \in \mathbb{R}_+^{N \times N}$ be the communication matrix between these devices describing the latency and bandwidth respectively, where the latency and bandwidth between device $d$ and $d'$ is $\alpha_{d,d'}$ and $\beta_{d,d'}$; $L$ denotes the total number of layers in the model to be served. We summarize all notations used in this paper in Appendix \ref{sec:notations} for easy reference.

\noindent \textbf{Formalization of the scheduling problem.}  
Given the above notations, we can formalize our scheduling problem as follows: suppose $\mathbf{d}_{i,j}$ is a subset of GPU devices that satisfies $\bigcup_{i,j} \mathbf{d}_{i,j} \subseteq \mathbf{D}$, 
and $\mathbf{d}_{i,j} \bigcap \mathbf{d}_{i',j'} = \emptyset, \quad \forall i \neq i' \vee j \neq j'$, where we suppose that the union of subsets of GPUs $\bigcup_{j} \mathbf{d}_{i,j}$ serves the $i$-th model replica as an independent pipeline, and the set of GPUs $\mathbf{d}_{i,j}$ serves the $j$-th stage in the $i$-th pipeline that holds $l_{i,j}$ transformer layers, if $\|\mathbf{d}_{i,j}\| > 1$, we indicate that the subset of GPUs $\mathbf{d}_{i,j}$ runs tensor model parallelism. An \textit{assignment} $\sigma$ is a mapping: $\mathbf{D} \rightarrow \left\{ \left(\mathbf{d}_{i,j}, l_{i,j}\right)\right\}$, corresponding to a layout of multiple inference worker groups to serve multiple replicas of the same model simultaneously. Consider a set of inference tasks $\mathbf{T}$ that satisfy some distribution\footnote{The most commonly-used case is Poisson distribution; while one can switch it for any particular distributions.} $\mathbf{T} \sim \mathcal{P}$. An \textit{optimal assignment} $\sigma^*$ can be defined as:
\begin{equation}
\begin{small}
\begin{aligned}
  \sigma^* = \arg\max_{\sigma \in \Sigma} & \quad \mathbb{E}_{\mathbf{T} \sim \mathcal{P}}\left[\textsc{SLO} \left(\textsc{C}_{\text{comm}}\left(\sigma\right) + \textsc{C}_{\text{comp}}\left(\sigma\right)\right) \right]\\
  s.t. & \quad \textsc{C}^{d}_{\text{mem}}\left(\sigma\right) \le M_d \quad \forall d \in \mathbf{D}.
\end{aligned}
\end{small}
\end{equation}
\noindent where $\textsc{C}_{\text{comm}}\left(\sigma\right)$,  $\textsc{C}_{\text{comp}}\left(\sigma\right)$,  $\textsc{C}_{\text{mem}}^{d}\left(\sigma\right)$ represents communication cost, computation cost, and memory limit for layout $\sigma$. 
Intuitively, the scheduling problem is to \textit{find an optimal assignment that partitions the device set to represent multiple independent inference pipeline groups that can maximize the inference service level objective (SLO) considering the computation cost, communication cost, and memory consumption constraints}. 
We summarize the estimation of the computation cost, communication cost, and memory consumption constraints in Table \ref{tab:formula}, and the detailed formulation is enumerated in Appendix \ref{sec:cost_model}. As one can see, solving this problem is obviously NP-hard; thus, we adopt a \textit{two-phase} search algorithm to tackle the problem:


\begin{itemize}[topsep=5pt, leftmargin=*]
    \vspace{-0.5em}
    \item We first generate a random partition of $\left\{\mathbf{d}_{i,\sim}\right\}$ of $\mathbf{D}$ where $\mathbf{d}_{i,\sim}$ is a set of GPU devices that can be leveraged to serve an independent pipeline group; given this partition, we determine the optimal layout of pipeline stage partition by an efficient dynamic programming method (Section \ref{sec:cost_estimate}). 
    
    \vspace{-0.5em}
    \item We propose an evolutionary algorithm that generates the random partition to be used in the first step, where we first define some hard constraints to reduce the search space of the random partition, and then we define some effective mutation operations that can be used to accelerate the convergence of the searching algorithm (Section \ref{sec:search}). 
\end{itemize}


\vspace{-0.5em}
\subsection{Optimizing the Layout of a Pipeline}
\vspace{-0.5em}
\label{sec:cost_estimate}

We first introduce how to solve a sub-optimization problem to find the optimal layout for an independent pipeline. Formally, given a set of GPUs $\mathbf{d}_{i,\sim}$, and a particular layer partition represented by $\{l_{i,j}, j=1,2,...,S_i\}$, this sub-optimization aims to find a local optimal assignment $\mathbf{d}_{i,\sim} \rightarrow \left\{\left(\mathbf{d}_{i,j}, l_{i,j}\right)\right\}$ that minimizes the end-to-end inference cost over the $i$-th pipeline defined in Equation \ref{eq:e2e_cost} below:
\begin{equation}
\label{eq:e2e_cost}
\begin{scriptsize}
\begin{aligned}
&\quad \quad \textsc{C}^{i}_{\text{comp}}\left(\left\{\mathbf{d}_{i,\sim}\right\}\right) + \textsc{C}^{i}_{\text{comm}}\left(\left\{\mathbf{d}_{i,\sim}\right\}\right) \\
 &= \sum_{j=1}^{S_i}{ \textsc{C}^{i,j}_{\text{comp}}\left(\mathbf{d}_{i,j}\right)} + 
\sum_{j=1}^{S_i}{\textsc{C}^{i,j}_{\text{comm-tp}}\left(\mathbf{d}_{i,j}\right)}+\sum_{j=1}^{S_i{-}1}{ \textsc{C}^{i,j}_{\text{comm-pp}}\left(\mathbf{d}_{i,j}\right)} \\
&s.t. \quad  \textsc{C}^{d}_{\text{mem}}\left(\left\{\mathbf{d}_{i,\sim}\right\}\right)  \le M_d \quad  \forall d \in \bigcup_{j} \mathbf{d}_{i,j}
\end{aligned}
\end{scriptsize}
\end{equation}
As one can notice, this sub-optimization problem is still NP-hard. To reduce the search space, we adopt the following heuristic: we force each tensor model parallel group to use the \textit{same type} of GPUs on the \textit{same} machine to avoid extensive cross-machine communication overhead. 
Following this heuristic, we use an additional notation to represent any GPU set by a vector $\boldsymbol{\tau} \in \mathcal{R}^{N_T}$, assuming that $N_T$ is the total number of different GPU types, and $\tau_k$ represents the number of the $k$-th type of GPU in this set. 

Next, we introduce the following dynamic programming (DP) algorithm to solve this problem. We define a memorization buffer $\textsc{DP}\in \mathcal{R}^{S_i \times \#_1 \times \#_2 ... \times \#_{N_T}}$, where $S_i$ is the total number of stages in this pipeline, $\#_k, k =1,...,N_T$ is the total number of $k$-th type of GPUs in the GPU set $\mathbf{d}_{i,\sim}$, the value 
$\textsc{DP}\left[j;\boldsymbol{\tau}\right]$ represents the communication and computation cost of assigning the first $j$ stage(s) in the GPU set represented by $\boldsymbol{\tau}$ if the memory limit for the first $j$ stage(s) is satisfied\footnote{$\textsc{DP}\left[j;\boldsymbol{\tau}\right] = +\infty$ if the memory limit is violated for any stage.}. For example, $\textsc{DP}\left[1;\left[2,0,...,0\right]\right]$ represents the communication and computation cost of assigning the first stage in two GPUs of the first type. If we initialize the values in $\textsc{DP}$ to $+\infty$ and set $\textsc{DP}\left[0; \boldsymbol{\tau}\right] = 0$, we can derive the transition formula as:
\begin{equation}
\begin{small}
\begin{aligned}
\textsc{DP}\left[j;\boldsymbol{\tau}\right] = \min_{\tau_k\cdot\mathbf{e}_k \subset \mathbf{d}_{i,\sim} } \left\{   \textsc{DP}\left[j-1; \boldsymbol{\tau} - \tau_k\cdot\mathbf{e}_k \right] + \right.\\ 
\left.  \textsc{C}^{i,j}_{\text{comp}}\left(\tau_k\cdot\mathbf{e}_k\right) + \textsc{C}^{i,j}_{\text{comm}}\left(\tau_k\cdot\mathbf{e}_k\right)  \right\}
\end{aligned}
\end{small}
\end{equation}
\noindent where $\tau_k\cdot\mathbf{e}_k$ is the vector representation of a set of $\tau_k$ $k$-th type GPUs, $\textsc{C}^{i,j}_{\text{comp}}\left(\tau_k\cdot\mathbf{e}_k\right)$ is defined by Table \ref{tab:formula}, and $\textsc{C}^{i,j}_{\text{comm}}\left(\tau_k\cdot\mathbf{e}_k\right)$ is the sum of tensor parallel communication cost and pipeline parallel communication cost follow the formulation in Table \ref{tab:formula}. Notice these two value will be set to $+\infty$ if the memory limit $\textsc{C}_{\text{mem}}^{d}\left(\tau_k\cdot\mathbf{e}_k\right)$ defined in Table \ref{tab:formula} is violated. Formally, we formalize a recursive implementation of the dynamic programming algorithm in Algorithm \ref{alg:dp} to estimate the minimal cost of this pipeline --- the assignment of the pipeline stage can be implemented by a standard back-track process over the memorization buffer $\textsc{DP}$ straightforwardly, which we do not enumerate here.

\noindent
\noindent \textbf{Complexity analysis}. The computation of the cost defined in Table \ref{tab:formula} can be done in constant time. For each stage, the proposed DP algorithm will visit at most $\prod_{k=1}^{N_T}{\#_k}$ different subsets of GPUs; the maximal depth of the recursion is $S_i$; thus, the total time complexity of the proposed DP algorithm is $\mathbb{O}\left(S_i\cdot\prod_{k=1}^{N_T}{\#_k}\right)$. This is much more efficient than the vanilla approach based on enumeration without the heuristic of forcing tensor parallelism to use the same type of GPUs --- in that case, each stage has to consider $2^{\left| \mathbf{d}_{i,~\sim}\right|}$ different subsets of GPUs, where the total time complexity grows to $\mathbb{O}\left(S_i\cdot 2^{\left| \mathbf{d}_{i,~\sim}\right|}\right)$. In practice, even a highly heterogeneous GPU pool probably only includes a limited variety of GPU types. The proposed DP algorithm can solve the sub-optimization problem efficiently. This process can be further accelerated by limiting the degree of tensor model parallelism to a smaller candidate set, e.g. $\{1,2,4,8\}$, which reduces the total time complexity to $\mathbb{O}\left(S_i\cdot 4^{N_T}\right)$.

\begin{algorithm}[tb]
\begin{small}
   \caption{Estimate optimal pipeline cost.}
   \label{alg:dp}
\begin{algorithmic}

\STATE {\bfseries Input:} Memorization Buffer $\textsc{DP}$, \\ \qquad \quad Current stage $j=1$, \\
\qquad \quad Assigned GPU set $\boldsymbol{\tau} = \emptyset$,\\
\qquad \quad Unassigned GPU set $\boldsymbol{\overline{\tau}} = \mathbf{d}_{i,\sim}$

\STATE \textbf{function} \textsc{Estimate-Pipeline-Cost}$\left(\textsc{DP},\: j, \:\boldsymbol{\tau}, \:\boldsymbol{\overline{\tau}} \right)$
    \STATE{\textcolor{green}{/* $N_T$ is number of different GPU types. */}}
     \FOR{$k=1$ {\bfseries to} $N_T$}
        \STATE{\textcolor{green}{/* $\#_k$ is number of type $k$-th GPU in $\boldsymbol{\overline{\tau}}$. */}}
        \FOR{$\tau_k=1$ {\bfseries to} $\#_k$}
            \STATE{\textcolor{green}{/* The current GPU set $\boldsymbol{\delta}$. */}}  
            \STATE{$\boldsymbol{\delta} \leftarrow \tau_k\cdot\mathbf{e}_k $;}
            \STATE{\textcolor{green}{/* Compute current cost $\textsc{c}$. */}}
            \STATE{$\textsc{c} \leftarrow \textsc{DP}\left[j{-}1; \boldsymbol{\tau}\right] + \textsc{C}^{i,j}_{\text{comp}}\left(\boldsymbol{\delta}\right) + \textsc{C}^{i,j}_{\text{comm}}\left(\boldsymbol{\delta}\right)$}
            \STATE{\textcolor{green}{/* Update global memorization buffer. */}}
            \IF{$\textsc{c} < \textsc{DP}\left[j;\boldsymbol{\tau}+\boldsymbol{\delta}\right]$}
                \STATE{$\textsc{DP}\left[j;\boldsymbol{\tau}+\boldsymbol{\delta}\right] \leftarrow \textsc{c}$;}
            \ENDIF
            \STATE{\textcolor{green}{/* Recursively assign the next stage(s). */}}
            \IF{ $j < S_i$ \textbf{and} $\boldsymbol{\overline{\tau}}{-}\boldsymbol{\delta} \ne \emptyset$ }
                \STATE{\textsc{Estimate-Pipeline-Cost}$\left(\textsc{DP},\: j{+}1, \:\boldsymbol{\tau}{+}\boldsymbol{\delta}, \:\boldsymbol{\overline{\tau}}{-}\boldsymbol{\delta} \right)$}
            \ENDIF
        \ENDFOR
    \ENDFOR
\STATE \textbf{end function}
\end{algorithmic}
\end{small}
\end{algorithm}

\vspace{-0.5em}
\subsection{Searching via Genetic Algorithm}
\vspace{-0.5em}
\label{sec:search}

Next, we introduce the genetic algorithm to solve the global optimization problem. Concretely, given the global set of GPUs $\mathbf{D}$, this genetic optimization problem finds an optimal partition of $\mathbf{D}$, to some independent pipeline groups $\mathbf{d}_{i,\sim}$ along with its stage partition $\left\{l_{i,1}, ..., l_{i,j}, l_{i,S_i}\right\}$. We enumerate the details of this genetic algorithm as follows. 

\noindent \textbf{Initialization}. A good initialization of the genetic algorithm is usually helpful in accelerating the whole search process. Thus, we initialize the population with a simple heuristic based on the communication condition. Formally, given the communication matrix $\mathbf{A}$ and $\mathbf{B}$, we execute the vanilla K-means algorithm to construct $M$ independent pipeline parallel groups, where the hyper-parameter $M$ is determined by the standard Elbow method. Intuitively, this helps the assignment avoid using slow cross-region communication links. Note that $M$ is not fixed after initialization, as we can change this by the mutation operations introduced next.

\noindent \textbf{Mutation operations.} We follow the vector representation of a set of GPUs introduced earlier in Section \ref{sec:cost_estimate} and define three mutation operations in the genetic algorithm. 

\begin{itemize}[topsep=5pt, leftmargin=*]
\vspace{-0.5em}
\item \underline{\texttt{Merge}}: Merge two pipeline groups into a single group. Formally, given two independent pipeline groups noted as $\boldsymbol{\tau}^1$ and $\boldsymbol{\tau}^2$, \textit{merge} is defined as: $\boldsymbol{\tau}^1, \boldsymbol{\tau}^2 \rightarrow \boldsymbol{\tau}$, where we have $\boldsymbol{\tau}= \boldsymbol{\tau}^1{+}\boldsymbol{\tau}^2$. 

\vspace{-0.5em}
\item \underline{\texttt{Split}}: Split a single pipeline group to two pipeline groups evenly. Given any independent pipeline group noted as $\boldsymbol{\tau}$, \textit{split} is defined as: $\boldsymbol{\tau} \rightarrow  \boldsymbol{\tau}^1,  \boldsymbol{\tau}^2$, where for any GPU type indexed by $k$, we have $\tau^1_k = \lfloor \frac{\tau_k}{2}\rfloor, \: \tau^2_k = \lceil \frac{\tau_k}{2}\rceil$.

\vspace{-0.5em}
\item \underline{\texttt{Swap}}: Move one GPU from one pipeline group to another pipeline group. Formally, given any two independent pipeline groups noted as $\boldsymbol{\tau}^1$ and $\boldsymbol{\tau}^2$, \textit{swap} is defined as: $\boldsymbol{\tau}^1, \boldsymbol{\tau}^2 \rightarrow \boldsymbol{\tau}'^1, \boldsymbol{\tau}'^2$, where for a particular sampled GPU type index $\widehat{k}$, we have\\
$\left\{\begin{array}{ll}
&\tau'^1_k = \tau^1_k, \:\:\: \tau'^2_k = \tau^2_k  \qquad \qquad\text{ if } k \ne \widehat{k};\\
&\tau'^1_k = \tau^1_k{+}1, \: \tau'^2_k = \tau^2_k{-}1 \: \qquad \text{otherwise.}
\end{array} \right.$
\vspace{-0.5em}
\end{itemize}

Notice that we also conduct some early checks for the off-springs generated by the above mutation operations. For example, after a \texttt{split} operation, if the total device memory of any of these two pipeline groups cannot even hold a copy of model parameters, we can simply remove this off-spring without running the dynamic programming algorithm (Algorithm \ref{alg:dp}) to accelerate search.    

\vspace{0.25em}
\noindent \textbf{Determine the pipeline partitions}. We adopt a simple expectation maximization algorithm to determine any particular pipeline stage partition (i.e. $\left\{l_{i,1}, ..., l_{i,j}, l_{i,S_i}\right\}$). Concretely, we first generate an even partition for any new off-spring generated by the mutation operations ($l_{i,j}=\frac{L}{S_i}, j=1,2,...,S_i$). Then after running the dynamic programming algorithm (Algorithm \ref{alg:dp}), we adjust the pipeline partition proportional to the total memory of the GPU set that currently serves this stage. We find that this heuristic is effective to find a good pipeline partition.   

\vspace{0.25em}
\noindent \textbf{Put it together}. We adopt a standard iterative procedure of the genetic algorithm: for each iteration, we conduct the mutation operation(s), run the dynamic algorithm to determine the optimal pipeline assignments, and adjust the pipeline partitions for the current off-springs. To estimate the expected SLO, we adopt the inference task simulator from AlpaServe~\cite{li2023alpaserve}.

\vspace{-0.5em}
\section{Evaluation}
\label{sec:eval}

\begin{figure*}[htbp]
    \centering
    \includegraphics[width=0.99\linewidth]{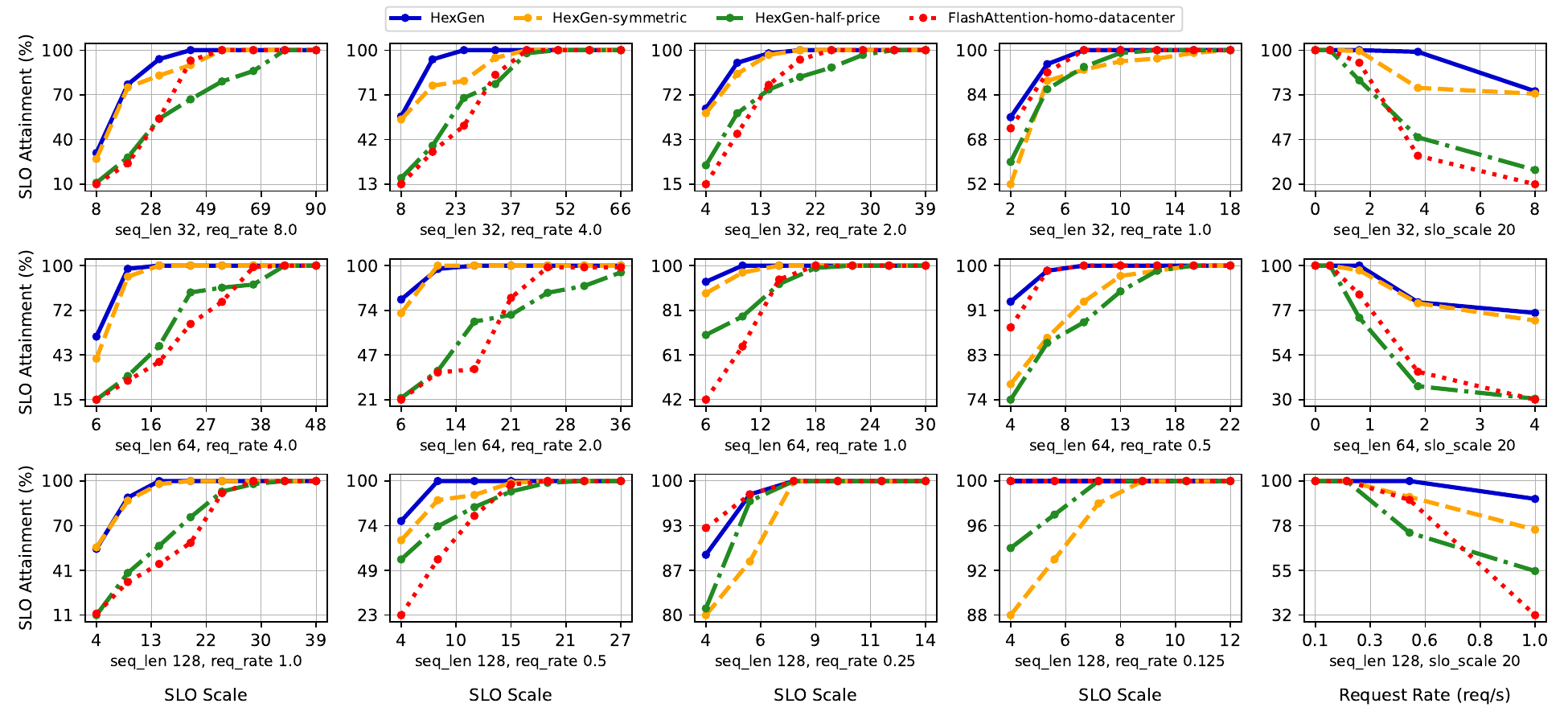}
    \vspace{-1em}
    \caption{SLO attainment results to evaluate cost performance trade-offs. Each row corresponds to a particular output sequence length (32, 64, 128). The first four columns correspond to different SLO scales ranging from 8 to 0.125 requests per second. The last column represents the performance comparison of various settings at different request rates.}
    \label{fig:slovssloscale}
    \vspace{-1.5em}
\end{figure*}

To evaluate the design and implementation of \sys, 
we ask the following essential questions:

\begin{itemize}[topsep=5pt, leftmargin=*]
    \vspace{-0.5em}
    \item \textit{What is the cost performance trade-off between the centralized homogeneous deployment and \sys over a heterogeneous decentralized GPU pool?}

    \vspace{-0.5em}
    \item \textit{What is the end-to-end performance comparison between \sys and the state-of-the-art centralized and decentralized generative inference systems?} 

    \vspace{-0.5em}
    \item \textit{How effective is the scheduling algorithm in finding the optimal assignment of the inference workflow?}
    \vspace{-0.5em}
\end{itemize}

\vspace{-0.5em}
\subsection{Experimental Setup}
\vspace{-0.5em}

\textbf{Runtime}. We perform evaluation in the following setups:

\begin{itemize}[topsep=5pt, leftmargin=*]
\vspace{-0.5em}
\item \underline{Homogeneous GPUs in a centralized data center.} We rent two \textit{AWS} on-demand \texttt{p4d.24xlarge} instances, each equipped with \texttt{8$\times$NVIDIA A100-40G} GPUs with a budget of $\$65.54/\text{hour}$ to represent the standard homogeneous case in a data center.

 \vspace{-0.5em}
\item \underline{{Heterogeneous GPUs across data centers.}} We rent GPUs from \textit{FluidStack}, a GPU cloud provider with services for various GPUs. Concretely, we considered two settings based on \textit{real availability}: i) \underline{heterogeneous-full-price}: we rent
two \texttt{3090Ti$\times$8} instances in Iceland, two \texttt{3090Ti$\times$3} instances in Norway, one \texttt{A5000$\times$8} in Nevada, two \texttt{A6000$\times$8} instances, one \texttt{A5000$\times 8$} instances and one \texttt{A40$\times$4} instances in Illinois with a budget of $\$65.04/\text{hour}$; ii) \underline{heterogeneous-half-price}: we rent two \texttt{3090Ti$\times$8} instances in Iceland, two \texttt{3090Ti$\times$3} instance in Norway, and one \texttt{A5000$\times$8} instance in Nevada with a budget of $\$29.6/\text{hour}$.\footnote{In the cross data center setting, we measure the network latency and bandwidth between GPUs in different regions by configuring a virtual private network through UDP hole punching, and benchmark the NCCL performance: the intra-region latency and bandwidth were $2$ ms and $5$ Gbps, while inter-region latency and bandwidth range from $40$ - $150$ ms and $0.3$ - $1.0$ Gbps.}
 \vspace{-0.5em}
\end{itemize}

\noindent \textbf{Baselines}. We carefully select state-of-the-art approaches as baselines. 
To understand the cost performance trade-offs and system efficiency of \sys, we compare: i) \sys under heterogeneous-full-price, ii) a truncated version of \sys without asymmetric parallel support under heterogeneous-full-price (the allocation of model replicas is still scheduled by our proposed search algorithm), iii) \sys under heterogeneous-half-price, and iv) \textsc{FlashAttention}~\cite{dao2023flashattention} under homogeneous-data-center setting.\footnote{Our implementation is identical to the standard \textsc{FlashAttention} implementation under a homogeneous setting.}    
To understand end-to-end performance, we compare \sys with \textsc{Huggingface-TGI}~\cite{huggingfaceTGI} as the state-of-the-art approach under the homogeneous setting and \textsc{Petals}~\cite{borzunov2023petals} as the state-of-the-art approach under decentralized heterogeneous setting.
To understand the efficiency of the proposed scheduling algorithm, we compare its convergence with a strawman policy based on random mutation.

\noindent \textbf{Evaluation metrics}.
\rebuttal{Following the generative inference evaluation setup from AlpaServe~\cite{li2023alpaserve}}, we test system performance based on SLO attainment\rebuttal{, refers to the proportion of requests that can be finished within a predefined performance threshold. Specifically, we measure SLO attainment in terms of the percentage of requests served within the time frame set by the SLO. We scale the SLO to various multiples of the execution latency of A100 GPUs (SLO Scale in \autoref{fig:slovssloscale}), which allows us to evaluate system performance under different levels of operational stringency.} We generate inference workload according to a Poisson process parameterized by the request rate. The consecutive requests (inter-arrival times) follow an exponential distribution. For a target SLO goal (e.g., $99\%$), we focus on two metrics: i) the minimum latency deadline required to achieve the desired attainment, and ii) the system's resilience to peak request rate. We apply the most popular open-source \textsc{Llama-2 (70B)} model on some real-world prompts~\cite{chatbotData}, and test output sequence lengths from $32$ to $128$, and request rates varying between $0.125$ - $10$ requests per second.

\vspace{-0.5em}
\subsection{Cost Performance Trade-off}
\vspace{-0.5em}

Figure \ref{fig:slovssloscale} illustrates a comprehensive comparison of the cost performance trade-off in terms of SLO attainment among \sys w/wo asymmetric parallel group support under the full budget in the heterogeneous setting, \sys under the half budget in the heterogeneous setting, and \textsc{FlashAttention} in the homogeneous setting. We want to highlight some interesting results. 

\noindent \textbf{Cost efficiency}. When given the relatively same budget, \sys under the full budget in the heterogeneous setting clearly outperforms \textsc{FlashAttention} in the homogeneous datacenter setting. In fact, \sys reaches up to $2.3\times$ and on average $1.5\times$ lower latency deadlines, and is capable of handling a peak request rate that is up to $4\times$ higher ($2\times$ higher on average). \rebuttal{Specifically, by analyzing the scheduling results\footnote{All the strategies chosen by the scheduling algorithm can be found in \autoref{sec:scheduling_results}.} for heterogeneous-full-price cases, we find that our scheduling approach always prioritizes intra-machine tensor model parallelism to minimize single request latency and employs inter-machine pipeline parallelism to reduce communication over limited bandwidth. It avoids cross-region communication due to ultra-low bandwidth and aims to maximize device memory utilization by incorporating as many model replicas as possible, thereby enhancing parallel request processing.} Furthermore, even when we reduce the budget in the heterogeneous setting by half, \sys still reveals similar performance to \textsc{FlashAttention} in the homogeneous setting. We believe that this is strong evidence to illustrate that a decentralized system such as \sys is capable of managing heterogeneous GPUs to provide more economical foundation model inference services without compromising service quality.  

\noindent \textbf{Asymmetric parallelism implementation.} We also conduct a group of benchmarks to compare \sys w/wo asymmetric parallel group support under the full budget in the heterogeneous setting. The experimental results reveal that the asymmetric parallelism implementation results in up to $1.8\times$ improvement in terms of reaching lower latency deadlines than the original symmetric implementation from \textsc{FlashAttention}. Additionally, the asymmetric parallelism implementation can manage up to $2\times$ higher peak traffic request rate ($1.5\times$ on average) compared to the symmetric counterpart. This indicates that besides effective scheduling, the asymmetric parallelism implementation is also necessary to unleash the potential of heterogeneous computational power.

\vspace{-0.5em}
\subsection{End-to-end System Performance}
\vspace{-0.5em}
\label{sec:eval_petals}

We compare the end-to-end system performance of \sys with state-of-the-art approaches for heterogeneous (\textsc{Petals}) and homogeneous (\textsc{Hugginface-TGI}) settings.

\begin{figure}[t]
    \centering
    \includegraphics[width=0.99\linewidth]{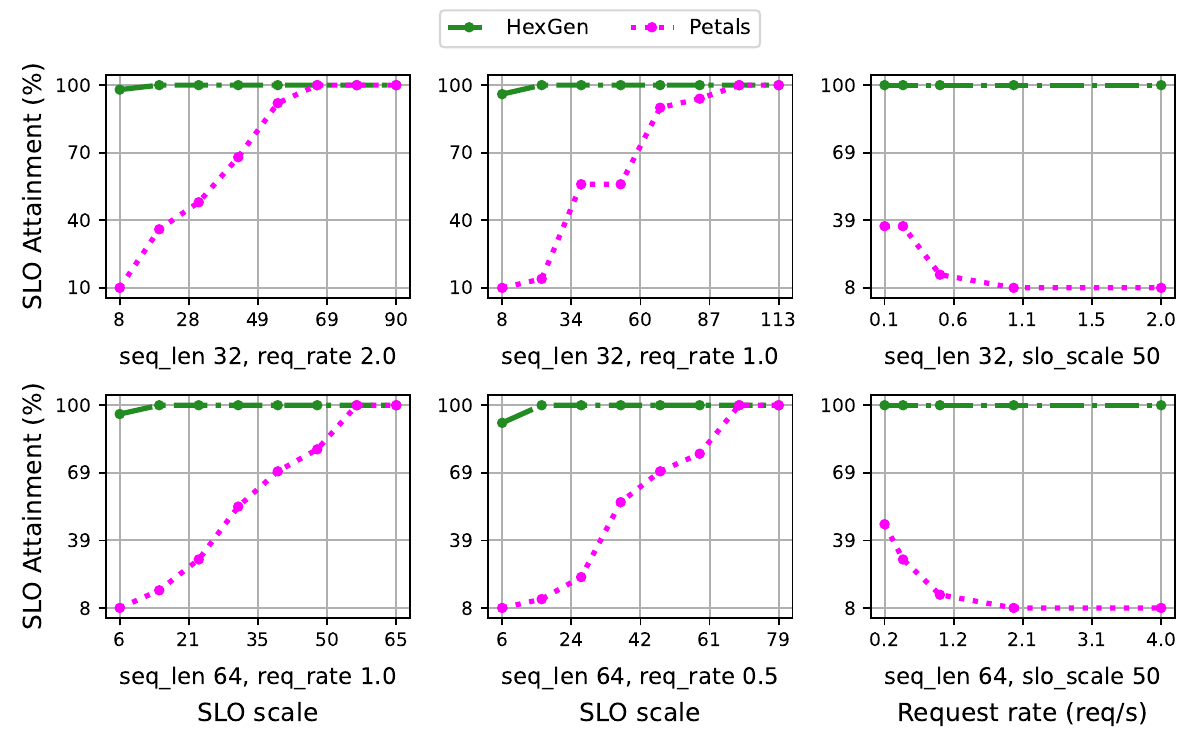}
    \caption{\sys and \textsc{Petals}. Two rows correspond to output sequence lengths of $32$ and $64$. First two columns illustrate the results of different SLO scales. The last column shows the effects of request rate on SLO attainment.}
    \label{fig:petals}
\end{figure}

\noindent\textbf{Compare with \textsc{Petals}}. We first compare \sys with \textsc{Petals}, the state-of-the-art decentralized inference service engine. Figure \ref{fig:petals} illustrates the comparison --- under the half-price budget in the heterogeneous setting, \sys outperforms \textsc{Petals} significantly by achieving up to $3.5\times$ lower latency deadline and managing to handle requests at up to $10\times$ higher rates. This is strong evidence to justify the design of \sys: \textsc{Petals} is mainly built on top of swarm parallelism~\cite{ryabinin2023swarm}, which heavily depends on dynamic adjustment of the collective learning paradigm to ensure the elasticity in a decentralized machine learning system; however, such a dynamic design compromises the inference service performance significantly when comparing with a system like \sys that is equipped with the careful design of static scheduling of the inference workflow.

\begin{figure}[t]
    \centering
    \includegraphics[width=0.99\linewidth]{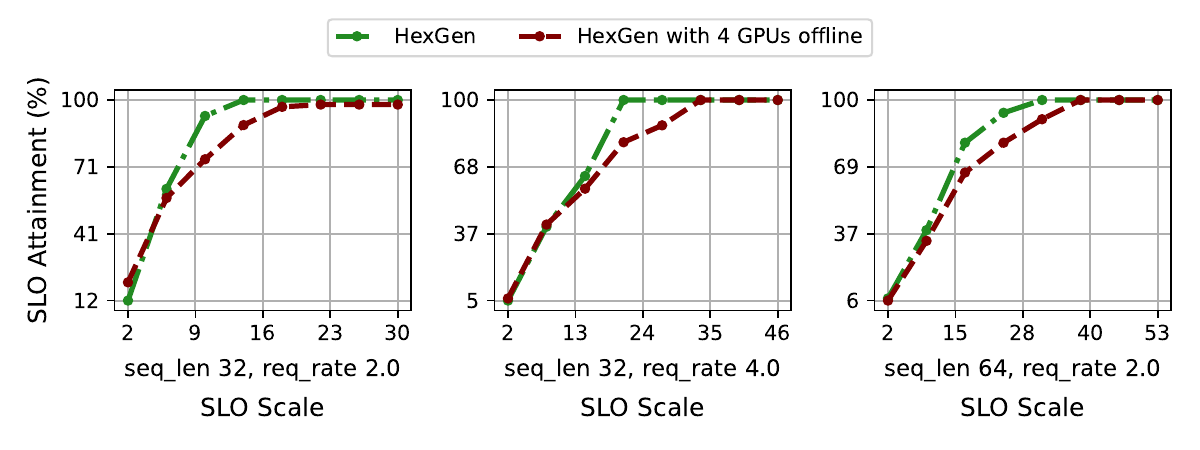}
    \caption{\rebuttal{SLO attainment results of \sys compared with \sys with 4 GPUs offline.}}
    \label{fig:offline}
\end{figure}

\rebuttal{
To evaluate \sys over dynamic GPU pools, we test a scenario where 4 GPUs leave the current allocation scheduled by \sys. In this case, \sys will re-run the search algorithm to find the new optimal allocation.  Interestingly, we find this simple policy is very effective---the genetic algorithm is based on local search, which demands much less iteration of searching when only a small portion of GPUs dynamically join or leave. 
We notice that \sys can rerun the searching algorithm in less than 30 seconds to find the new optimal allocation. Figure \ref{fig:offline} illustrates \sys's performance before and after the 4 GPUs become offline.  We see the performance gap is considerably small under such dynamics. In addition, we find that the performance of \sys with 4 GPUs offline is still significantly better than \textsc{Petals}.}

\begin{figure}[t]
    \centering
    \includegraphics[width=0.99\linewidth]{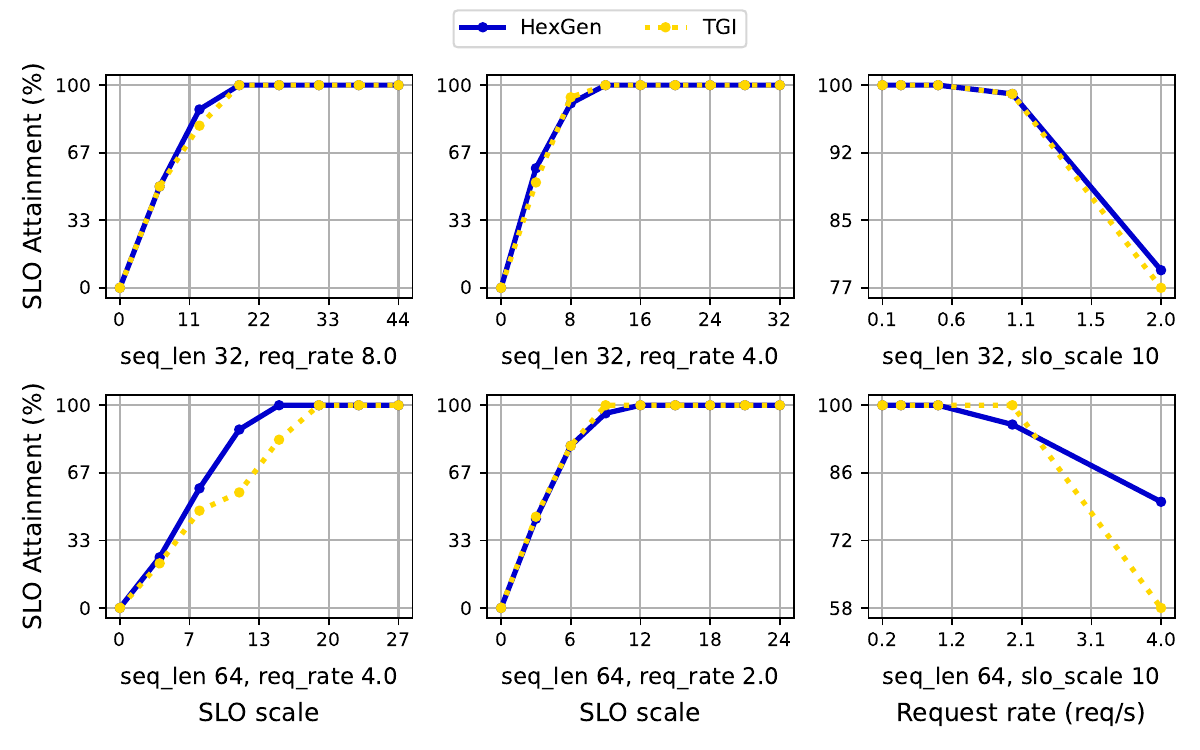}
    \caption{\sys v.s. \textsc{Huggingface-TGI}. Two rows represent output sequence lengths of $32$ and $64$. First two columns show the result of different SLO scales. The last column shows the effects of request rate on SLO attainment.}
    \label{fig:tgi}
    \vspace{-1.5em}
\end{figure}

\noindent\textbf{Compare with \textsc{Huggingface-TGI}}. We also compare \sys under the full budget in heterogeneous setting with \textsc{Huggingface-TGI} in the homogeneous data center settings. In this case, \sys gets almost the same end-to-end performance in both latency and request handling, achieves up to $1.25\times$ lower latency deadlines, and is able to handle requests at the same level of rates.

\vspace{-0.5em}
\subsection{Effectiveness of the Scheduling Algorithm}
\vspace{-0.5em}

\begin{figure}[htbp]
    \centering
    \includegraphics[width=0.7\linewidth]{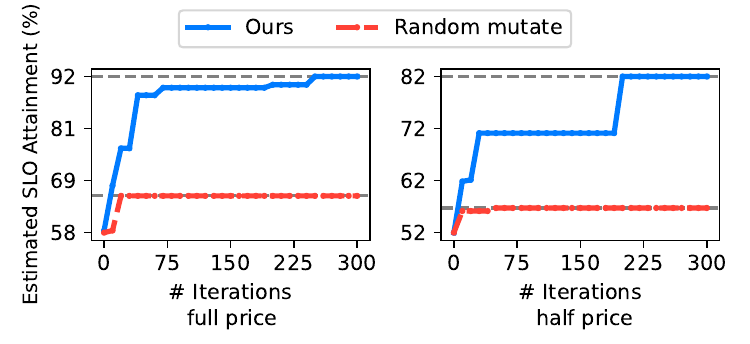}
    \vspace{-1.0em}
    \caption{Convergence comparison of the proposed search strategy and random mutation.}
    \label{fig:mutateslo}
    \vspace{-1.0em}
\end{figure}

\begin{figure}[t]
    \centering
    \includegraphics[width=0.99\linewidth]{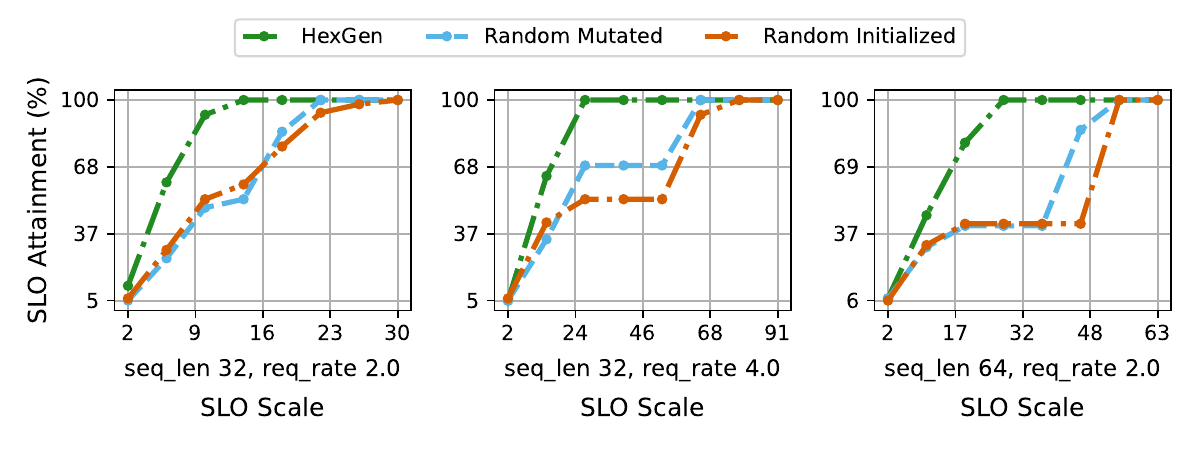}
    \caption{\rebuttal{SLO attainment results of \sys compared with random mutation based results and results w/o any searching efforts.}}
    \label{fig:random}
    \vspace{-2.0em}
\end{figure}

To evaluate the effectiveness of the scheduling algorithm, we disable the advanced mutation strategies introduced in Section \ref{sec:search} by replacing this part with random mutation and comparing the convergence behavior with the proposed algorithm.  We benchmarked the full-price and half-price cases, where we set the output sequence length to $32$, and the SLO scale as $5$. Figure \ref{fig:mutateslo} illustrates the result. \rebuttal{The proposed search strategy takes 2.1 and 1.5 minutes to identify the optimal assignments for full-price and half-price heterogeneous scenarios. This search process runs once before the system is initially deployed, which makes its time cost negligible.}
We see that our proposed constrained mutation policy significantly outperforms random mutation --- the proposed strategy can find assignments that can manage around $26\%$ more SLO attainments than random mutation and converges much faster, while random mutation gets stuck in some local minimum. 
Additionally, we verified that in both cases, the estimated SLO attainments ($92\%$ and $82\%$) closely align with the actual attainments ($94\%$ and $86\%$) demonstrated in \autoref{fig:slovssloscale}.  \rebuttal{We also provide a direct performance comparison between random initialized allocation (the allocation after executing the initialization method mentioned in \autoref{sec:search}), random mutated policy through evolution and \sys's result in the heterogeneous-half-price scenario. The results are illustrated in Figure \ref{fig:random}.}

\vspace{-0.5em}
\section{Related Work}
\label{sec:rel}

\noindent \textbf{Foundation model inference optimization}. There have been many efforts to accelerate the inference service in terms of both system optimization and algorithm design. 
On the system side, research efforts have focused on enhancing hardware efficiency through meticulous system optimizations~\cite{fang2021turbotransformers,yu2022orca,li2023alpaserve,kwon2023efficient,dao2022flashattention}.
On the algorithm side, some advanced algorithm designs have also been proposed~\cite{leviathan2023fast,yao2022zeroquant,liu2023deja}, including speculative decoding~\cite{spector2023accelerating}, multiple-head decoding mechanism~\cite{medusa} and low precision computation such as quantization~\cite{yao2022zeroquant,frantar2022gptq,xiao2022smoothquant,lin2023awq}, sparsification~\cite{frantar2023sparsegpt,liu2023deja} and distillation~\cite{kwon2022alphatuning}.

\noindent \textbf{Decentralized computation platform}.
Recently, there have been some emerging research attempts on deploying machine learning computations across a variety of decentralized and heterogeneous computational resources~\cite{ali2022optimizing,miao2023sdpipe,zhang2023efficient,stoica2021cloud,bhat2023sakshi}, e.g., distributed training in a collaborative environment~\cite{diskin2021distributed, yuan2022decentralized,ryabinin2023swarm}. However, most of this work does not focus on the system implementation and scheduling for generative inference workflows. Perhaps the most relevant effort is Petals~\cite{borzunov2022petals}, which allows users to donate different GPUs to perform inference and small-scale fine-tuning. However, Petals is mainly based on dynamic coordination from swarm parallelism~\cite{ryabinin2023swarm}, whose performance is limited by the lack of scheduling of the decentralized inference. 


\vspace{-0.5em}
\section{Conclusion}
\vspace{-0.5em}
\label{sec:conclude}
In this paper, we explore the opportunity to deploy the inference service of foundation models via a heterogeneous regime with devices of different computation capacities connected over a heterogeneous network. Toward this end, we propose \sys, a generative inference framework with asymmetric parallel support and an effective scheduling algorithm to accommodate such deployment. Our empirical study suggests that when given the same budget, \sys can outperform the centralized homogeneous deployment by either achieving $2.3\times$ lower latency deadlines or tolerating up to $4\times$ more traffic request rate; additionally, \sys also significantly outperforms Petals, the state-of-the-art decentralized collaborative inference prototype by $10\times$ more traffic request rate. We will make the \sys system fully open-sourced and hope that such a system can contribute to the democratization of the usage of foundation models.


\section*{Impact Statements}

This paper presents work whose goal is to advance the field of Machine Learning. There are many potential societal consequences of our work, none of which we feel must be specifically highlighted here.

\nocite{langley00}

\bibliography{example_paper}

\begin{thebibliography}{45}
\providecommand{\natexlab}[1]{#1}
\providecommand{\url}[1]{\texttt{#1}}
\expandafter\ifx\csname urlstyle\endcsname\relax
  \providecommand{\doi}[1]{doi: #1}\else
  \providecommand{\doi}{doi: \begingroup \urlstyle{rm}\Url}\fi

\bibitem[Ali et~al.(2022)Ali, Pinciroli, Yan, and Smirni]{ali2022optimizing}
Ali, A., Pinciroli, R., Yan, F., and Smirni, E.
\newblock Optimizing inference serving on serverless platforms.
\newblock \emph{Proceedings of the VLDB Endowment}, 15\penalty0 (10):\penalty0 2071--2084, 2022.

\bibitem[Athlur et~al.(2022)Athlur, Saran, Sivathanu, Ramjee, and Kwatra]{athlur2022varuna}
Athlur, S., Saran, N., Sivathanu, M., Ramjee, R., and Kwatra, N.
\newblock Varuna: scalable, low-cost training of massive deep learning models.
\newblock In \emph{Proceedings of the Seventeenth European Conference on Computer Systems}, pp.\  472--487, 2022.

\bibitem[Bhat et~al.(2023)Bhat, Chen, Cheng, Fang, Hebbar, Kannan, Rana, Sheng, Tyagi, Viswanath, et~al.]{bhat2023sakshi}
Bhat, S., Chen, C., Cheng, Z., Fang, Z., Hebbar, A., Kannan, S., Rana, R., Sheng, P., Tyagi, H., Viswanath, P., et~al.
\newblock Sakshi: Decentralized ai platforms.
\newblock \emph{arXiv preprint arXiv:2307.16562}, 2023.

\bibitem[Bommasani et~al.(2021)Bommasani, Hudson, Adeli, Altman, Arora, von Arx, Bernstein, Bohg, Bosselut, Brunskill, et~al.]{bommasani2021opportunities}
Bommasani, R., Hudson, D.~A., Adeli, E., Altman, R., Arora, S., von Arx, S., Bernstein, M.~S., Bohg, J., Bosselut, A., Brunskill, E., et~al.
\newblock On the opportunities and risks of foundation models.
\newblock \emph{arXiv preprint arXiv:2108.07258}, 2021.

\bibitem[Borzunov et~al.(2022)Borzunov, Baranchuk, Dettmers, Ryabinin, Belkada, Chumachenko, Samygin, and Raffel]{borzunov2022petals}
Borzunov, A., Baranchuk, D., Dettmers, T., Ryabinin, M., Belkada, Y., Chumachenko, A., Samygin, P., and Raffel, C.
\newblock Petals: Collaborative inference and fine-tuning of large models.
\newblock \emph{arXiv preprint arXiv:2209.01188}, 2022.
\newblock URL \url{https://arxiv.org/abs/2209.01188}.

\bibitem[Borzunov et~al.(2023)Borzunov, Baranchuk, Dettmers, Riabinin, Belkada, Chumachenko, Samygin, and Raffel]{borzunov2023petals}
Borzunov, A., Baranchuk, D., Dettmers, T., Riabinin, M., Belkada, Y., Chumachenko, A., Samygin, P., and Raffel, C.
\newblock Petals: Collaborative inference and fine-tuning of large models.
\newblock In \emph{Proceedings of the 61st Annual Meeting of the Association for Computational Linguistics (Volume 3: System Demonstrations)}, pp.\  558--568, 2023.

\bibitem[Bubeck et~al.(2023)Bubeck, Chandrasekaran, Eldan, Gehrke, Horvitz, Kamar, Lee, Lee, Li, Lundberg, et~al.]{bubeck2023sparks}
Bubeck, S., Chandrasekaran, V., Eldan, R., Gehrke, J., Horvitz, E., Kamar, E., Lee, P., Lee, Y.~T., Li, Y., Lundberg, S., et~al.
\newblock Sparks of artificial general intelligence: Early experiments with gpt-4.
\newblock \emph{arXiv preprint arXiv:2303.12712}, 2023.

\bibitem[Cai et~al.(2023)Cai, Li, Geng, Peng, and Dao]{medusa}
Cai, T., Li, Y., Geng, Z., Peng, H., and Dao, T.
\newblock Medusa: Simple framework for accelerating llm generation with multiple decoding heads.
\newblock \url{https://github.com/FasterDecoding/Medusa}, 2023.

\bibitem[Cheatham et~al.(1996)Cheatham, Fahmy, Stefanescu, and Valiant]{cheatham1996bulk}
Cheatham, T., Fahmy, A., Stefanescu, D., and Valiant, L.
\newblock Bulk synchronous parallel computing—a paradigm for transportable software.
\newblock \emph{Tools and Environments for Parallel and Distributed Systems}, pp.\  61--76, 1996.

\bibitem[Dao(2023)]{dao2023flashattention}
Dao, T.
\newblock Flashattention-2: Faster attention with better parallelism and work partitioning.
\newblock \emph{arXiv preprint arXiv:2307.08691}, 2023.

\bibitem[Dao et~al.(2022)Dao, Fu, Ermon, Rudra, and R{\'e}]{dao2022flashattention}
Dao, T., Fu, D., Ermon, S., Rudra, A., and R{\'e}, C.
\newblock Flashattention: Fast and memory-efficient exact attention with io-awareness.
\newblock \emph{Advances in Neural Information Processing Systems}, 35:\penalty0 16344--16359, 2022.

\bibitem[Diskin et~al.(2021)Diskin, Bukhtiyarov, Ryabinin, Saulnier, Sinitsin, Popov, Pyrkin, Kashirin, Borzunov, Villanova~del Moral, et~al.]{diskin2021distributed}
Diskin, M., Bukhtiyarov, A., Ryabinin, M., Saulnier, L., Sinitsin, A., Popov, D., Pyrkin, D.~V., Kashirin, M., Borzunov, A., Villanova~del Moral, A., et~al.
\newblock Distributed deep learning in open collaborations.
\newblock \emph{Advances in Neural Information Processing Systems}, 34:\penalty0 7879--7897, 2021.

\bibitem[Fang et~al.(2021)Fang, Yu, Zhao, and Zhou]{fang2021turbotransformers}
Fang, J., Yu, Y., Zhao, C., and Zhou, J.
\newblock Turbotransformers: an efficient gpu serving system for transformer models.
\newblock In \emph{Proceedings of the 26th ACM SIGPLAN Symposium on Principles and Practice of Parallel Programming}, pp.\  389--402, 2021.

\bibitem[Frantar \& Alistarh(2023)Frantar and Alistarh]{frantar2023sparsegpt}
Frantar, E. and Alistarh, D.
\newblock Sparsegpt: Massive language models can be accurately pruned in one-shot, 2023.

\bibitem[Frantar et~al.(2022)Frantar, Ashkboos, Hoefler, and Alistarh]{frantar2022gptq}
Frantar, E., Ashkboos, S., Hoefler, T., and Alistarh, D.
\newblock Gptq: Accurate post-training quantization for generative pre-trained transformers.
\newblock \emph{arXiv preprint arXiv:2210.17323}, 2022.

\bibitem[Guo et~al.(2022)Guo, Guo, Kim, Hildred, and Daudjee]{guo2022hydrozoa}
Guo, R., Guo, V., Kim, A., Hildred, J., and Daudjee, K.
\newblock Hydrozoa: Dynamic hybrid-parallel dnn training on serverless containers.
\newblock \emph{Proceedings of Machine Learning and Systems}, 4:\penalty0 779--794, 2022.

\bibitem[Huang et~al.(2019)Huang, Cheng, Bapna, Firat, Chen, Chen, Lee, Ngiam, Le, Wu, et~al.]{huang2019gpipe}
Huang, Y., Cheng, Y., Bapna, A., Firat, O., Chen, D., Chen, M., Lee, H., Ngiam, J., Le, Q.~V., Wu, Y., et~al.
\newblock Gpipe: Efficient training of giant neural networks using pipeline parallelism.
\newblock \emph{Advances in neural information processing systems}, 32, 2019.

\bibitem[HuggingFace(2022)]{huggingfaceAccelerate}
HuggingFace.
\newblock Hugging face accelerate.
\newblock \url{https://huggingface.co/docs/accelerate/index}, 2022.

\bibitem[HuggingFace(2023)]{huggingfaceTGI}
HuggingFace.
\newblock Text generation inference.
\newblock \url{https://huggingface.co/docs/text-generation-inference/index}, 2023.

\bibitem[Institute(2023)]{falcon180b}
Institute, T.~I.
\newblock Falcon 180b, 2023.
\newblock URL \url{https://falconllm.tii.ae/falcon-180b.html}.

\bibitem[Kwon et~al.(2022)Kwon, Kim, Bae, Yoo, Kim, Park, Kim, Ha, Sung, and Lee]{kwon2022alphatuning}
Kwon, S.~J., Kim, J., Bae, J., Yoo, K.~M., Kim, J.-H., Park, B., Kim, B., Ha, J.-W., Sung, N., and Lee, D.
\newblock Alphatuning: Quantization-aware parameter-efficient adaptation of large-scale pre-trained language models.
\newblock \emph{arXiv preprint arXiv:2210.03858}, 2022.

\bibitem[Kwon et~al.(2023)Kwon, Li, Zhuang, Sheng, Zheng, Yu, Gonzalez, Zhang, and Stoica]{kwon2023efficient}
Kwon, W., Li, Z., Zhuang, S., Sheng, Y., Zheng, L., Yu, C.~H., Gonzalez, J., Zhang, H., and Stoica, I.
\newblock Efficient memory management for large language model serving with pagedattention.
\newblock In \emph{Proceedings of the 29th Symposium on Operating Systems Principles}, pp.\  611--626, 2023.

\bibitem[Leviathan et~al.(2023)Leviathan, Kalman, and Matias]{leviathan2023fast}
Leviathan, Y., Kalman, M., and Matias, Y.
\newblock Fast inference from transformers via speculative decoding.
\newblock In \emph{International Conference on Machine Learning}, pp.\  19274--19286. PMLR, 2023.

\bibitem[Li et~al.(2023)Li, Zheng, Zhong, Liu, Sheng, Jin, Huang, Chen, Zhang, Gonzalez, et~al.]{li2023alpaserve}
Li, Z., Zheng, L., Zhong, Y., Liu, V., Sheng, Y., Jin, X., Huang, Y., Chen, Z., Zhang, H., Gonzalez, J.~E., et~al.
\newblock $\{$AlpaServe$\}$: Statistical multiplexing with model parallelism for deep learning serving.
\newblock In \emph{17th USENIX Symposium on Operating Systems Design and Implementation (OSDI 23)}, pp.\  663--679, 2023.

\bibitem[LibP2P(2023)]{libp2p}
LibP2P.
\newblock A modular network stack, 2023.
\newblock URL \url{https://libp2p.io/}.

\bibitem[Lin et~al.(2023)Lin, Tang, Tang, Yang, Dang, and Han]{lin2023awq}
Lin, J., Tang, J., Tang, H., Yang, S., Dang, X., and Han, S.
\newblock Awq: Activation-aware weight quantization for llm compression and acceleration.
\newblock \emph{arXiv preprint arXiv:2306.00978}, 2023.

\bibitem[Liu et~al.(2023)Liu, Wang, Dao, Zhou, Yuan, Song, Shrivastava, Zhang, Tian, Re, et~al.]{liu2023deja}
Liu, Z., Wang, J., Dao, T., Zhou, T., Yuan, B., Song, Z., Shrivastava, A., Zhang, C., Tian, Y., Re, C., et~al.
\newblock Deja vu: Contextual sparsity for efficient llms at inference time.
\newblock In \emph{International Conference on Machine Learning}, pp.\  22137--22176. PMLR, 2023.

\bibitem[Lmsys(2023)]{chatbotData}
Lmsys.
\newblock Chatbot arena conversations.
\newblock \url{https://huggingface.co/datasets/lmsys/chatbot_arena_conversations}, 2023.

\bibitem[Miao et~al.(2023{\natexlab{a}})Miao, Oliaro, Zhang, Cheng, Wang, Wong, Chen, Arfeen, Abhyankar, and Jia]{miao2023specinfer}
Miao, X., Oliaro, G., Zhang, Z., Cheng, X., Wang, Z., Wong, R. Y.~Y., Chen, Z., Arfeen, D., Abhyankar, R., and Jia, Z.
\newblock Specinfer: Accelerating generative llm serving with speculative inference and token tree verification.
\newblock \emph{arXiv preprint arXiv:2305.09781}, 2023{\natexlab{a}}.

\bibitem[Miao et~al.(2023{\natexlab{b}})Miao, Shi, Yang, Cui, and Jia]{miao2023sdpipe}
Miao, X., Shi, Y., Yang, Z., Cui, B., and Jia, Z.
\newblock Sdpipe: A semi-decentralized framework for heterogeneity-aware pipeline-parallel training.
\newblock \emph{Proceedings of the VLDB Endowment}, 16\penalty0 (9):\penalty0 2354--2363, 2023{\natexlab{b}}.

\bibitem[Narayanan et~al.(2019)Narayanan, Harlap, Phanishayee, Seshadri, Devanur, Ganger, Gibbons, and Zaharia]{narayanan2019pipedream}
Narayanan, D., Harlap, A., Phanishayee, A., Seshadri, V., Devanur, N.~R., Ganger, G.~R., Gibbons, P.~B., and Zaharia, M.
\newblock Pipedream: generalized pipeline parallelism for dnn training.
\newblock In \emph{Proceedings of the 27th ACM Symposium on Operating Systems Principles}, pp.\  1--15, 2019.

\bibitem[Narayanan et~al.(2021)Narayanan, Shoeybi, Casper, LeGresley, Patwary, Korthikanti, Vainbrand, Kashinkunti, Bernauer, Catanzaro, et~al.]{narayanan2021efficient}
Narayanan, D., Shoeybi, M., Casper, J., LeGresley, P., Patwary, M., Korthikanti, V., Vainbrand, D., Kashinkunti, P., Bernauer, J., Catanzaro, B., et~al.
\newblock Efficient large-scale language model training on gpu clusters using megatron-lm.
\newblock In \emph{Proceedings of the International Conference for High Performance Computing, Networking, Storage and Analysis}, pp.\  1--15, 2021.

\bibitem[NVIDIA(2022)]{fastertransformer}
NVIDIA.
\newblock Fastertransformer.
\newblock \url{https://github.com/NVIDIA/FasterTransformer}, 2022.

\bibitem[Ryabinin et~al.(2023)Ryabinin, Dettmers, Diskin, and Borzunov]{ryabinin2023swarm}
Ryabinin, M., Dettmers, T., Diskin, M., and Borzunov, A.
\newblock Swarm parallelism: Training large models can be surprisingly communication-efficient.
\newblock \emph{arXiv preprint arXiv:2301.11913}, 2023.

\bibitem[Spector \& Re(2023)Spector and Re]{spector2023accelerating}
Spector, B.~F. and Re, C.
\newblock Accelerating llm inference with staged speculative decoding.
\newblock In \emph{Workshop on Efficient Systems for Foundation Models@ ICML2023}, 2023.

\bibitem[Stoica \& Shenker(2021)Stoica and Shenker]{stoica2021cloud}
Stoica, I. and Shenker, S.
\newblock From cloud computing to sky computing.
\newblock In \emph{Proceedings of the Workshop on Hot Topics in Operating Systems}, pp.\  26--32, 2021.

\bibitem[Thorpe et~al.(2023)Thorpe, Zhao, Eyolfson, Qiao, Jia, Zhang, Netravali, and Xu]{thorpe2023bamboo}
Thorpe, J., Zhao, P., Eyolfson, J., Qiao, Y., Jia, Z., Zhang, M., Netravali, R., and Xu, G.~H.
\newblock Bamboo: Making preemptible instances resilient for affordable training of large $\{$DNNs$\}$.
\newblock In \emph{20th USENIX Symposium on Networked Systems Design and Implementation (NSDI 23)}, pp.\  497--513, 2023.

\bibitem[Touvron et~al.(2023)Touvron, Martin, Stone, Albert, Almahairi, Babaei, Bashlykov, Batra, Bhargava, Bhosale, et~al.]{touvron2023llama}
Touvron, H., Martin, L., Stone, K., Albert, P., Almahairi, A., Babaei, Y., Bashlykov, N., Batra, S., Bhargava, P., Bhosale, S., et~al.
\newblock Llama 2: Open foundation and fine-tuned chat models.
\newblock \emph{arXiv preprint arXiv:2307.09288}, 2023.

\bibitem[Xiao et~al.(2022)Xiao, Lin, Seznec, Demouth, and Han]{xiao2022smoothquant}
Xiao, G., Lin, J., Seznec, M., Demouth, J., and Han, S.
\newblock Smoothquant: Accurate and efficient post-training quantization for large language models.
\newblock \emph{arXiv preprint arXiv:2211.10438}, 2022.

\bibitem[Yang et~al.(2023)Yang, Wu, Luo, Chiang, Bhardwaj, Kwon, Zhuang, Luan, Mittal, Shenker, et~al.]{yang2023skypilot}
Yang, Z., Wu, Z., Luo, M., Chiang, W.-L., Bhardwaj, R., Kwon, W., Zhuang, S., Luan, F.~S., Mittal, G., Shenker, S., et~al.
\newblock $\{$SkyPilot$\}$: An intercloud broker for sky computing.
\newblock In \emph{20th USENIX Symposium on Networked Systems Design and Implementation (NSDI 23)}, pp.\  437--455, 2023.

\bibitem[Yao(2023)]{yao2023open}
Yao, X.
\newblock {Open Compute Framework: Peer-to-Peer Task Queue for Foundation Model Inference Serving}, September 2023.
\newblock URL \url{https://github.com/autoai-org/OpenComputeFramework}.

\bibitem[Yao et~al.(2022)Yao, Aminabadi, Zhang, Wu, Li, and He]{yao2022zeroquant}
Yao, Z., Aminabadi, R.~Y., Zhang, M., Wu, X., Li, C., and He, Y.
\newblock Zeroquant: Efficient and affordable post-training quantization for large-scale transformers.
\newblock \emph{arXiv preprint arXiv:2206.01861}, 2022.

\bibitem[Yu et~al.(2022)Yu, Jeong, Kim, Kim, and Chun]{yu2022orca}
Yu, G.-I., Jeong, J.~S., Kim, G.-W., Kim, S., and Chun, B.-G.
\newblock Orca: A distributed serving system for $\{$Transformer-Based$\}$ generative models.
\newblock In \emph{16th USENIX Symposium on Operating Systems Design and Implementation (OSDI 22)}, pp.\  521--538, 2022.

\bibitem[Yuan et~al.(2022)Yuan, He, Davis, Zhang, Dao, Chen, Liang, Re, and Zhang]{yuan2022decentralized}
Yuan, B., He, Y., Davis, J., Zhang, T., Dao, T., Chen, B., Liang, P.~S., Re, C., and Zhang, C.
\newblock Decentralized training of foundation models in heterogeneous environments.
\newblock \emph{Advances in Neural Information Processing Systems}, 35:\penalty0 25464--25477, 2022.

\bibitem[Zhang et~al.(2023)Zhang, Li, Zhao, Xu, Lu, Xiao, Han, Yang, and Du]{zhang2023efficient}
Zhang, Q., Li, J., Zhao, H., Xu, Q., Lu, W., Xiao, J., Han, F., Yang, C., and Du, X.
\newblock Efficient distributed transaction processing in heterogeneous networks.
\newblock \emph{Proceedings of the VLDB Endowment}, 16\penalty0 (6):\penalty0 1372--1385, 2023.

\end{thebibliography}
\bibliographystyle{icml2024}

\newpage
\appendix
\onecolumn
\textbf{Contents:} In Section \ref{sec:notations}, we summarize the notations throughout this paper for easy reference. In Section \ref{sec:cost_model}, We enumerate how we formulate the generative inference cost in terms of computation time and communication time in the tensor model and pipeline parallelism and the corresponding memory limit. In Section \ref{sec:task}, we present an extended discussion on \sys system implementation about the task coordinator. In Section \ref{sec:batching}, we discuss the limitation of batching implementation for \sys. In Section \ref{sec:extended_rel}, we elaborate on the extended version of related works.

\section{Summarization of Notations}
\label{sec:notations}

We summarize the notations used for the scheduling algorithm of this paper in Table \ref{sample-table} for easy reference. 



\begin{table*}[h!]
\caption{Table of notations.} 
\label{sample-table}
\vspace{-1em}
\begin{center}
\begin{footnotesize}
\begin{tabular}{c | l}
\hline
\textbf{Symbol} & \textbf{Description} \\
\hline
$N$  &  Total number of GPUs to serve generative inference. \\
$N_T$  &  Total number of distinct GPU types. \\
$\mathbf{D}$  &  Set of $N$ GPU devices. \\
$M_d$      & Memory limit of device $d$. \\
$m_d$      & GPU memory bandwidth of device $d$. \\
$c_d$      & Tensor core computation power of device $d$. \\
$\mathbf{A}$  &  Communication matrix between devices describing the latency. \\
$\mathbf{B}$  &  Communication matrix between devices describing the bandwidth. \\
$\alpha_{d,d'}$  & Latency between devices $d$ and $d'$. \\
$\beta_{d,d'}$   & Bandwidth between devices $d$ and $d'$. \\
$L$  &  Total number of layers in the model to be served. \\
$H$ & Size of the hidden dimension in a transformer block. \\
$B_{\text{type}}$  & Byte size for computational precision. \\
$s^{\text{in}}_t$    & Length of input sequence for task $t$. \\
$s^{\text{out}}_t$   & Length of output sequence for task $t$. \\
$b_t$   & Batch size allocated for inference task $t$. \\
$\mathbf{d}_{i,j}$   & Set of GPUs serves the $j$-th stage in the $i$-th pipeline. \\
$l_{i,j}$   & Number of transformer layers in the $j$-th stage of the $i$-th pipeline. \\
$\sigma$  & Mapping of models to devices. \\
$\mathbf{T}$  &  Set of inference tasks. \\
$\mathcal{P}$  &  Distribution of inference tasks. \\
$S_i$   &  Number of stages in the $i$-th pipeline. \\
$k$  &  Identifier for a GPU type. \\
$\#_k$  & Total number of $k$-th type of GPUs in the GPU set $\mathbf{d}_{i,\sim}$. \\
$\boldsymbol{\tau}$  &  Vector denoting any GPU set. \\
$\tau_k$  &  Number of the $k$-th type of GPU in $\boldsymbol{\tau}$. \\
$\tau_k\cdot\mathbf{e}_k$  &  Vector denoting a set of $\tau_k$ $k$-th type GPUs, where $\mathbf{e}_k$ denotes $k$-th standard basis vector. \\
$M$  &  Number of independent pipeline parallel groups. \\
\hline
\end{tabular}
\end{footnotesize}
\end{center}
\end{table*}

\section{Modeling the Generative Inference Cost}
\label{sec:cost_model}

We enumerate how we formulate the generative inference cost in terms of computation time and communication time in the tensor model and pipeline parallelism and the corresponding memory limit.  
Following the notation introduced in Sections \ref{sec:preliminary} and \ref{sec: problem}. Let $B_{\text{type}}$ be the number of bytes for the precision of inference computation, e.g., $B_{\text{type}}\left(\textsc{fp16}\right)=2$; for a particular inference task $t \in \mathbf{T}$, where $b_{t}$ is the batch size, $s^{\text{in}}_{t}$ be the sequence length of input prompt and $s^{\text{out}}_{t}$ be the sequence length of output token generation. Comprehensively, given a particular assignment $\sigma$ output noted as $\left\{ \mathbf{d}_{i,j} \right\}$, we estimate the communication cost, the computation cost, and the memory limit as follows: 

\noindent \textbf{Model the computation time}. Recall the introduction of inference computation in Section \ref{sec:preliminary}; one can notice that most of the computation is spent on matrix multiplications in a transformer block, while only a very small portion of computation is spent on other components, such as no-linear activation functions, batch normalization, etc. To simplify the modeling, we model the computation time that mainly comes from two sources: (i) \textit{scanning the model parameters through GPU high memory bandwidths to tensor cores}, since usually $b_t \ll H$, we ignore the time to scan the intermediate computation results; (ii) \textit{the computation w.r.t the matrix multiplications in the transformer block}, here we assume the computation time is only determined by the total float number operations in the transformer block and the devices' peak FLOPS, ignores other potential dynamic factors that can influence the execution time. Given a particular layer running tensor model parallelism over a set of GPU devices noted as $\mathbf{d}_{i,j}$ ($j$-th stage in $i$-th pipeline), the computation time of $l_{i,j}$ layer of transformers noted as $\textsc{C}^{i,j}_{\text{comp}}\left(\mathbf{d}_{i,j}\right)$ can be determined by the following formula: 
%
\begin{equation}
\label{eq:comp_i_j}
\begin{aligned}
\textsc{C}^{i,j}_{\text{comp}}\left(\mathbf{d}_{i,j}\right)
=  \max_{d \in \mathbf{d}_{i,j}}\left( \frac{12H^2 B_{\text{type}}s_t^{\text{out}}}{\left|\mathbf{d}_{i,j}\right| m_d} \right) \cdot l_{i,j}
 +  \max_{d \in \mathbf{d}_{i,j}}\left( \frac{24 b_t \left(s^{\text{in}}_t + s^{\text{out}}_t\right) H^2}{\left|\mathbf{d}_{i,j}\right| c_d}  \right) \cdot l_{i,j} 
\end{aligned}
\end{equation}

%
\noindent where the first part is to estimate the memory scan cost --- the model parameter has to be scanned $s_t^{\text{out}}$ times for all the generated tokens; $12H^2$ represents the total number of parameters in a transformer layer (e..g, the $\kappa$-th layer includes the weights $\wk^\kappa, \wq^\kappa, \wv^\kappa, \wo^\kappa \in \mathcal{R}^{H \times H} $, $\wa^\kappa \in \mathcal{R}^{H \times 4H}$, and $\wb^\kappa \in \mathcal{R}^{4H \times H}$); the second part is about the matrix multiplication --- for each layer, there are $\frac{24 b_t s^{\text{in}}_t H^2}{\left|\mathbf{d}_{i,j}\right| c_d}$ float point operations for the prefill phase and $\frac{24 b_t H^2}{\left|\mathbf{d}_{i,j}\right| c_d}$ float point operations for each token. To be more specific, the total number of float point operations for a matrix multiplication between matrix $\mathbf{X} \in \mathcal{R}^{H_1 \times H_2}$ and matrix $\mathbf{Y} \in \mathcal{R}^{H_2 \times H_3}$ can be calculated by $2H_1H_2H_3$ --- $\frac{24 b_t s^{\text{in}}_t H^2}{\left|\mathbf{d}_{i,j}\right| c_d}$ and $\frac{24 b_t H^2}{\left|\mathbf{d}_{i,j}\right| c_d}$ are calculated by accumulating the float point operations of matrix multiplications in the prefill phase and for one generated token in the decoding phrase as we introduced in Section \ref{sec:preliminary}. Notice that if $\| \mathbf{d}_{i,j} \| = 1$, Equation \ref{eq:comp_i_j} still holds, which represents only one GPU serves this stage without running tensor model parallelism.

\noindent \textbf{Model the communication time}. We make the following assumptions about modeling the communication cost: (i) for the point-to-point communication, we use the $\alpha - \beta$ cost model, where the communication time can be estimated by $\alpha+\frac{B}{\beta}$, and $B$ is the number of bytes that needs to be communicated; (ii) for collective communications, we adopt the bulk synchronous parallel (BSP) model~\cite{cheatham1996bulk} --- the communication execution is subdivided into supersteps, each associated with a global synchronization; and the total cost of a superstep is the \textit{max} over all processors at that superstep. 

We first model the communication in \textbf{tensor model parallelism}. In tensor model parallelism, each GPU needs to conduct two \texttt{AllReduce} operations for each transformer block. By the BSP model, we assume the tensor that needs to be aggregated by \texttt{AllReduce} will be partitioned to equal chunks, and accomplished by two supersteps (phases) (\texttt{ReduceScatter} and \texttt{AllGather}), in the \texttt{ReduceScatter} phase, each GPU sends its chunk to every other GPU and conduct the aggregation for its chunk; in the \texttt{AllGather} phase, each GPU sends its aggregated chunk to every other GPU. Formally, given a particular layer running tensor model parallelism over a set of GPU devices noted as $\mathbf{d}_{i,j}$ ($j$-th stage in $i$-th pipeline), the tensor model parallel communication time of $l_i,j$ layer of transformers noted as $\textsc{C}^{i,j}_{\text{comm-tp}}\left(\mathbf{d}_{i,j}\right)$ can be determined by the following formula: 
\begin{equation}
\label{eq:comm_tp}
\begin{aligned}
\textsc{C}^{i,j}_{\text{comm-tp}}\left(\mathbf{d}_{i,j}\right) 
= & \max_{d \in \mathbf{d}_{i,j}}\left( \sum_{d' \in \mathbf{d}_{i,j} - \{d\}} \left(\alpha_{d, d'} + \frac{b_t s^{\text{in}}_{t} H B_{\text{type}} } { {\left|\mathbf{d}_{i,j}\right| \beta}_{d, d'}}\right) \right) \cdot 4 l_{i,j}\\
 + & \max_{d \in \mathbf{d}_{i,j}} \left( \sum_{d' \in \mathbf{d}_{i,j} - \{d\}} \left(\alpha_{d, d'} + \frac{b_t H B_{\text{type}} } { {\left|\mathbf{d}_{i,j}\right| \beta}_{d, d'}}\right)  \right) \cdot 4 s_{t}^{\text{out}} l_{i,j} 
\end{aligned}
\end{equation}
\noindent Where the first part is to estimate the tensor model parallel communication cost in prefill phase, while the second term corresponds to the decoding phrase. Notice that one can verify that if $\|\mathbf{d}_{i,j}\|=1$, $\textsc{C}^{i,j}_{\text{comm-tp}}\left(\mathbf{d}_{i,j}\right) = 0$, which illustrate that there is no communication cost if only one GPU is serving this stage. 

Next, we model the communication cost in \textbf{pipeline parallelism}. To estimate the communication cost between nearby stages ($\mathbf{d}_{i,j}$ and $ \mathbf{d}_{i,j+1}$) in pipeline parallelism, we model this communication process by using the fastest link between these two stages. Formally, the pipeline parallel communication cost noted as $\textsc{C}^{i,j}_{\text{comm-pp}}\left(\mathbf{d}_{i,j}\right)$ can be formalized as:
\begin{equation}
\label{eq:comm_pp}
\begin{aligned}
\textsc{C}^{i,j}_{\text{comm-pp}}\left(\mathbf{d}_{i,j}\right)
= \min_{d \in \mathbf{d}_{i,j}, d' \in \mathbf{d}_{i,j+1}}\left( \alpha_{d,d'} + \frac{b_t s^{\text{in}}_{t} H B_{\text{type}}}{\beta_{d,d'}} \right) + \min_{d \in \mathbf{d}_{i,j}, d' \in \mathbf{d}_{i,j+1}}\left( \alpha_{d,d'} + \frac{b_t H B_{\text{type}}}{\beta_{d,d'}} \right) \cdot s_{t}^{\text{out}} 
\end{aligned}
\end{equation}
\noindent Where the first part is to estimate the pipeline model parallel communication cost in prefill phase, while the second term corresponds to the decoding phrase.

\noindent \textbf{Model the memory constraint}.
The GPU memory footprint mainly comes from three sources during inference computation: (i) to store the model parameters; (ii) to store the intermediate results including \texttt{key} and \texttt{value} for \textit{each} transformer blocks; and (iii) some activation caches---in an efficient implementation, such memory buffer will be reused during the computation for all transformer blocks, in our implementation the number of such buffer is $4$. Formally the memory constraint $\textsc{C}^{d}_{\text{mem}}\left(\mathbf{d}_{i,j}\right)$ for a device $d \in \mathbf{d}_{i,j}$ can be formalized as:
\begin{equation}
\label{eq:mem}
\begin{split}
\textsc{C}^{d}_{\text{mem}}\left(\mathbf{d}_{i,j}\right) 
= \left(\frac{12H^2 B_{\text{type}}}{\left|\mathbf{d}_{i,j}\right|} + \frac{2 b_t \left(s^{\text{in}}_t + s^{\text{out}}_t\right) H B_{\text{type}}}{\left|\mathbf{d}_{i,j}\right|} \right) \times l_{i,j} 
+ 4 b_t \left(s^{\text{in}}_t + s^{\text{out}}_t\right) H B_{\text{type}}
\end{split}
\end{equation}


\rebuttal{
\noindent \textbf{Performance alignment.} To evaluate the accuracy of our cost model, we conduct a suite of micro-benchmarks to validate the output of our cost estimation by comparing it with the actual execution time. The results are listed in Table \ref{tab:align} and suggest that our cost model can align with the actual execution accurately. 
}

\begin{table}[ht]
\centering
\caption{\rebuttal{Comparison of Benchmarked and Estimated Performance Metrics.}}
\rebuttal{
\begin{footnotesize}
\begin{tabular}{l | l | c | c | c | c}
\hline
\begin{tabular}[c]{@{}l@{}}\textbf{Input/Output}\\ \textbf{Length}\end{tabular} & \begin{tabular}[c]{@{}l@{}}\textbf{Parallel}\\ \textbf{Configuration}\end{tabular} & \begin{tabular}[c]{@{}l@{}}\textbf{Prefill Time -}\\ \textbf{Benchmarked}\end{tabular} & \begin{tabular}[c]{@{}l@{}}\textbf{Prefill Time -}\\ \textbf{Estimated}\end{tabular} & \begin{tabular}[c]{@{}l@{}}\textbf{Decode Time -}\\ \textbf{Benchmarked}\end{tabular} & \begin{tabular}[c]{@{}l@{}}\textbf{Decode Time -}\\ \textbf{Estimated}\end{tabular} \\
\hline
256/32 & TP=8 & 2.72s & 2.99s & 2.43s & 2.46s \\
       & TP=4 PP=2 & 3.79s & 3.85s & 2.25s & 2.14s \\
       & TP=2 PP=4 & 5.26s & 5.25s & 3.29s & 3.04s \\
       & PP=8 & 8.04s & 7.83s & 6.04s & 5.60s \\
512/64 & TP=8 & 3.04s & 3.10s & 4.76s & 4.92s \\
       & TP=4 PP=2 & 4.16s & 4.10s & 4.32s & 4.28s \\
       & TP=2 PP=4 & 5.57s & 5.63s & 6.65s & 6.08s \\
       & PP=8 & 8.27s & 8.49s & 12.4s & 11.2s \\
\hline
\end{tabular}
\end{footnotesize}
}
\label{tab:align}
\end{table}

\section{Task Coordinator}
\label{sec:task}

To deploy \sys in a real heterogeneous decentralized environment, we implement a  
\textbf{task coordinator}. The task coordinator manages the GPUs from the heterogeneous pool and organizes the independent pipeline parallel worker groups according to the optimal allocation based on the approach we introduced in Section \ref{sec:schedule}. Concretely, the task coordinator is mainly based on an open-source implementation of decentralized computation coordination~\cite{yao2023open} that utilizes libP2P~\cite{libp2p} to establish connections among the work groups in a peer-to-peer network. 
When an inference request is received by the task coordinator, the request will be directed to an appropriate worker group according to the scheduling result. 

\section{Limitation of Batching Implementation}
\label{sec:batching}
When we compare \sys with \textsc{Huggingface-TGI}, 
we realize that one particular reason that actually limits \sys from reaching its full potential when compared with \textsc{Huggingface-TGI} --- state-of-the-art foundation model inference services usually include some advanced batching policy to improve the hardware efficiency significantly; the current version of \sys has not integrated this feature yet since the batching policy under the heterogeneous setting issues some unique challenges given the diversified executing time and memory limit of each independent pipeline groups, which makes the current batching policy difficult to integrate. 
We acknowledge this limitation in the current version of \sys and leave this as an important future work to further improve \sys's end-to-end inference performance under the heterogeneous setting.

\section{Extended Related Works}
\label{sec:extended_rel}

\noindent \textbf{Foundation model inference optimization}. There have been many efforts to accelerate the inference service in terms of both system optimization and algorithm design. 
On the system side, research efforts have focused on enhancing hardware efficiency through meticulous system optimizations~\cite{fang2021turbotransformers,yu2022orca,li2023alpaserve,kwon2023efficient,dao2022flashattention}. For example, AlpaServe~\cite{li2023alpaserve} proposes a concrete analysis of model parallel strategies and model placement to improve inference service efficiency; 
PagedAttention~\cite{kwon2023efficient} introduces an advanced memory management system to batch inference jobs inspired by the classic design of virtual memory and paging; FLashAttention~\cite{dao2023flashattention} leveraged GPU memory hierarchy to reduce GPU memory access significantly, leading to runtime speedup. 
On the algorithm side, some advanced algorithm designs have also been proposed~\cite{leviathan2023fast,yao2022zeroquant,liu2023deja}. For example, speculative decoding~\cite{leviathan2023fast,miao2023specinfer,spector2023accelerating} based algorithms improve the system efficiency by leveraging a small approximation model for prediction and the original model for parallel verification. A similar idea has been explored by Medusa~\cite{medusa}, which implements a multiple-head decoding mechanism for parallel token verification. Low precision computation has also been explored to speed up generative inference, such as quantization~\cite{yao2022zeroquant,frantar2022gptq,xiao2022smoothquant,lin2023awq}, sparsification~\cite{frantar2023sparsegpt,liu2023deja} and distillation~\cite{kwon2022alphatuning}.

\noindent \textbf{Decentralized computation platform}.
Recently, there have been some emerging research attempts on deploying machine learning computations across a variety of decentralized and heterogeneous computational resources~\cite{ali2022optimizing,miao2023sdpipe,zhang2023efficient}. For example, sky computing~\cite{stoica2021cloud, yang2023skypilot} builds an additional layer above classic cloud platforms to access more economic computation resources and enable interoperability between multi-clouds; SAKSHI~\cite{bhat2023sakshi} has proposed a blueprint to advocate for the hosting and delivery of reliable AI models in the landscape of AI services. There have also been some attempts to deploy machine learning training in a collaborative environment~\cite{diskin2021distributed, yuan2022decentralized,ryabinin2023swarm}. However, most of this work does not focus on the system implementation and scheduling for generative inference workflows. Perhaps the most relevant effort is Petals~\cite{borzunov2022petals}, which allows users to donate different GPUs to perform inference and small-scale fine-tuning. However, Petals is mainly based on dynamic coordination from swarm parallelism~\cite{ryabinin2023swarm}, whose performance is limited by the lack of scheduling of the decentralized inference as we illustrated in Section \ref{sec:eval_petals}.




\rebuttal{
\section{Case Study of the Scheduling Result}
\label{sec:scheduling_results}
We list the partition results generated by \sys in the heterogeneous-full-price scenario. We use the following representation to describe the scheduled results. We use an array to specify one independent inference pipeline, and the number represents the degree of tensor parallelism for this pipeline stage. For example, [4,2] indicates a two-stage pipeline, where the first stage has a tensor parallel degree of 4, and the second stage has a tensor parallel degree of 2. The scheduled results are listed in \autoref{tab:gpu_strategy}.
}

\rebuttal{
\noindent \textbf{Partition breakdown.} In Iceland, two 8 $\times$ 3090Ti instances deploy [4,4] strategy to support two model replicas, each function within a single machine. Two 3 $\times$ 3090Ti instances in Norway employ a [2,1,1,2] strategy; cross-machine pipeline parallelism communication happens between the 2nd and 3rd stage. In Nevada, 8 $\times$ A5000 GPUs deploy a [4,4] strategy. In Illinois, 12 $\times$ A6000 GPUs serve four model replicas,  each with strategy [2,1]. The remaining 4 $\times$ A6000 / 4 $\times$ A40 is split into four groups, each 2 $\times$ A6000 / 2 $\times$ A40 and another 2 $\times$ A5000 deploys a [2,2] strategy, inter-machine communication is finished by pipeline parallelism.
}

\begin{table}[ht]
\centering
\caption{\rebuttal{GPU Deployment and Strategy by Region.}}
\label{tab:gpu_strategy}
\rebuttal{
\begin{tabular}{l | l | l}
\hline
\textbf{Region} & \textbf{GPU Configuration} & \textbf{Strategy} \\
\hline
Iceland & $8 \times 3090\text{Ti}$ & $[4,4]$ \\
 & $8 \times 3090\text{Ti}$ & $[4,4]$ \\
Norway & $3 \times 3090\text{Ti} + 3 \times 3090\text{Ti}$ & $[2,1,1,2]$ \\
Nevada & $8 \times \text{A5000}$ & $[4,4]$ \\
Illinois & $3 \times \text{A6000}$ & $[2,1]$ \\
 & $3 \times \text{A6000}$ & $[2,1]$ \\
 & $3 \times \text{A6000}$ & $[2,1]$ \\
 & $3 \times \text{A6000}$ & $[2,1]$ \\
 & $2 \times \text{A6000} + 2 \times \text{A5000}$ & $[2,2]$ \\
  & $2 \times \text{A6000} + 2 \times \text{A5000}$ & $[2,2]$ \\
   & $2 \times \text{A40} + 2 \times \text{A5000}$ & $[2,2]$ \\
    & $2 \times \text{A40} + 2 \times \text{A5000}$ & $[2,2]$ \\
\hline
\end{tabular}
}
\end{table}

\rebuttal{
\noindent \textbf{Interesting insight.} In a homogeneous setting, the 16 A100 GPUs can serve 4 \textsc{Llama-2 (70B)} model replicas. While in a heterogeneous setting, the 58 cloud GPUs with various types can serve a maximum of 12 \textsc{Llama-2 (70B)} model replicas with various hybrid parallel configurations within the same budget. Here are some interesting insights:
\begin{itemize}
    \vspace{-1.em}
    \item In this scenario, although individual inference tasks in a heterogeneous environment may experience increased latency, the overall performance of the system significantly improves.
    \vspace{-0.5em}
    \item Comprehensively, our scheduling approach prioritizes intra-machine tensor model parallelism to minimize single request latency and employs inter-machine pipeline parallelism to reduce communication over limited bandwidth. It avoids cross-region communication due to ultra-low bandwidth and aims to maximize device memory utilization by incorporating as many model replicas as possible.
    \vspace{-0.5em}
    \item Additionally, asymmetric parallelism plays a significant role in enhancing system performance, primarily by allowing the adoption of more adaptable parallel strategies to minimize extensive communication via low-bandwidth links. For instance, consider a scenario with 4 A5000 GPUs on one machine and 2 on another. Given the exceptionally low intercommunication bandwidth and the requirement for at least 6×24 GB GPUs to support a 70B model replica, the optimal configuration involves establishing a tensor model parallel group of 4 as the first pipeline stage and a group of 2 as the second. This setup is preferred because pipeline parallel communication demands are substantially lower than those of tensor model parallelism.
\end{itemize}
}



\end{document}